\documentclass[graybox]{svmult}

\usepackage[authoryear,round]{natbib}
\usepackage{type1cm}        
\usepackage{makeidx}         
\usepackage{multicol}        
\usepackage[bottom]{footmisc}

\usepackage{newtxtext}       %
\usepackage[varvw]{newtxmath}       

\usepackage{graphicx}
\usepackage{epstopdf}
\usepackage{algorithm}
\usepackage{hyperref}
\newcommand{\bq}{\begin{eqnarray*}}
\newcommand{\eq}{\end{eqnarray*}}
\newcommand{\bqn}{\begin{eqnarray}}
\newcommand{\eqn}{\end{eqnarray}}

\makeindex             

\begin{document}

\title*{Topological  Embedding of Human Brain Networks with Applications to Dynamics of Temporal Lobe Epilepsy}
\titlerunning{Topological  Embedding of Human Brain Networks} 
\author{Moo K. Chung, Ji Bi Che, Veena A. Nair, Camille Garcia Ramos, Jedidiah Ray Mathis, Vivek Prabhakaran, Elizabeth Meyerand, Bruce P. Hermann, Jeffrey R. Binder, Aaron F. Struck}
\authorrunning{Chung et al.} 
\institute{
Moo K. Chung \at Department of Biostatistics and Medical Informatics, University of Wisconsin, Madison, WI, USA \email{mkchung@wisc.edu}
\and
Ji Bi Che \at OpenAI, San Francisco, USA \email{jibiche1994@gmail.com}
\and Veena A. Nair \at Department of Radiology, University of Wisconsin-Madison, USA  \email{vnair@uwhealth.org}
\and Camille Garcia Ramos \at Department of  Neurology, University of Wisconsin-Madison, USA \email{garciaramos@wisc.edu}
\and Jedidiah Ray Mathis \at Department of Neurology,  Medical College of Wisconsin, USA \email{jmathis@mcw.edu}
\and Vivek Prabhakaran \at  Department of Radiology, University of Wisconsin-Madison, USA \email{prabhakaran@wisc.edu}
\and Elizabeth Meyerand \at Departments of Medical Physics \& Biomedical Engineering, University of Wisconsin-Madison, USA  \email{memeyerand@wisc.edu}
\and Bruce P. Hermann \at Department of Neurology, University of Wisconsin-Madison, USA \email{hermann@neurology.wisc.edu}
\and Jeffrey R. Binder \at Department of Neurology,  Medical College of Wisconsin, USA \email{jbinder@mcw.edu}
\and Aaron F. Struck \at Department of Neurology, University of Wisconsin-Madison, USA \email{struck@neurology.wisc.edu}
}

%
%
\maketitle

\abstract{
We introduce a novel, data-driven topological data analysis (TDA) approach for embedding brain networks into a lower-dimensional space in quantifying the dynamics of temporal lobe epilepsy (TLE) obtained from resting-state functional magnetic resonance imaging (rs-fMRI). This embedding facilitates the orthogonal projection of 0D and 1D topological features, allowing for the visualization and modeling of the dynamics of functional human brain networks in a resting state. We then quantify the topological disparities between networks to determine the coordinates for embedding. This framework enables us to conduct a coherent statistical inference within the embedded space. Our results indicate that brain network topology in TLE patients exhibits increased rigidity in 0D topology but more rapid flections compared to that of normal controls in 1D topology.}

\section{Introduction}

The analysis of brain networks through graph theory has become a cornerstone in understanding the intricate neural connections that underpin human cognition and behavior \citep{bassett.2017,sporns.2003,vanwijk.2010,chung.2017.BC}. Traditionally, graph-theoretical analyses of brain networks have relied on metrics such as node degree, clustering coefficients, and path lengths, extracted from adjacency matrices representing neural connections. However, these analyses often hinge on a predetermined threshold to binarize weighted networks, introducing a level of arbitrariness that can skew the final statistical results and their interpretation \citep{lee.2011.tmi,chung.2013.MICCAI}. The demand for a more robust analytical framework that transcends the limitations of threshold-dependent network analysis has led to the integration of  persistent homology \citep{edelsbrunner.2010}, a key concept from topological data analysis (TDA), into  brain network analysis \citep{lee.2011.MICCAI, lee.2012.tmi, petri.2014, sizemore.2018, sizemore.2019,vaccarino.2022}.

Persistent homology offers a multiscale perspective, unveiling the topological features of brain networks across a continuum of thresholds, and thus avoid the arbitrariness \citet{lee.2011.MICCAI, lee.2012.tmi}.  This approach not only captures the essence of complex neural architectures but also provides a stable representation that is less susceptible to the noise and variability inherent in neuroimaging data. By tracing the persistence of topological features—such as connected components and loops—across different scales, persistent homology encapsulates the hierarchical organization of brain connectivity, offering a nuanced understanding of its topological characterization of brain networks \citet{chung.2017.CNI,chung.2019.ISBI, kuang.2019,yoo.2017}. The adaptability of persistent homology in scrutinizing brain  networks has been underscored in many recent studies. \citet{sizemore.2019} and \citet{xing.2022} showcased the utility of persistent homology for assessing time-dependent shifts in the topological attributes of networks. \citet{aktas.2019} leveraged persistent homology to monitor the progression of network cliques, while \citet{billings.2021} explored its application in representing brain networks through simplicial complexes. Persistent homology's effectiveness in mapping out the spatial organization of cliques and cycles in brain networks was examined by \citet{sizemore.2018}. In functional brain connectivity analysis using EEG data, persistent homology's relevance was demonstrated by \citet{khalid.2014,caputi.2021}. \citet{chung.2023.NI} applied persistent homology to investigate the structural covariate networks associated with abnormal white matter in maltreated children. These contributions highlight persistent homology's capacity to offer a comprehensive framework for analyzing networks across multiple scales.

Although numerous studies have applied persistent homology to static networks or summarized time-varying networks in a static manner, the exploration of dynamic patterns in persistent homology for evolving brain networks remains less common, with only a handful of exceptions \citep{yoo.2016, santos.2019,song.2020.ISBI, giusti.2016, sizemore.2018,chung.2024.PLOS}. These time-varying networks, represented as a series of graphs, encapsulate the fluctuating landscape of neural interactions. The challenge lies in deciphering the underlying topological patterns that characterize these dynamic changes \citep{chung.2024.PLOS}. This paper presents a novel data embedding technique called the {\em Topological Phase Diagram}, which embeds time-varying brain networks into a 2D space to visually illustrate the temporal evolution of brain networks, providing a clear and accessible means for modeling and interpreting dynamic neural interactions. Our framework employs the Wasserstein distance between persistent diagrams to generate a 2D topological profile of the data. Traditionally, the Wasserstein distance, or Kantorovich–Rubinstein metric, defined for probability distributions, faced scalability issues \citep{shi.2016,su.2015,ma.2023,chung.2023.NI,vallender.1974,canas.2012,berwald.2018}. By establishing a direct correlation between the Wasserstein distance and network edge weights, our method enhances scalability and adaptability. We achieve a computational complexity of $\mathcal{O}(p \log p)$ for most network manipulation tasks like matching and averaging. This efficiency enables the execution of demanding procedures, such as topological embedding and clustering, with relative ease.

The dynamics of brain networks play a critical role in understanding cognitive functions, including reasoning, attention, and executive functions.  \citet{haier.1988} investigated the correlation between cortical glucose metabolism and cognitive functions using positron emission tomography (PET). The study found significant associations between glucose metabolism in specific cortical areas and performance on cognitive tasks, suggesting that higher metabolic rates are linked with enhanced reasoning and attention capabilities, providing insights into how brain energy consumption correlates with cognitive performance and intelligence. \citet{eisenberg.2005} found that educational experience correlates positively with glucose metabolism in brain regions crucial for sustained attention and learning, suggesting that educational experience may enhance metabolic efficiency in key brain regions, thereby potentially contributing to improvements in the g-factor and overall general intelligence.  \citet{pittau.2012}  investigated the patterns of altered functional connectivity in patients with mesial temporal lobe epilepsy (MTLE). Using fMRI, the connectivity patterns between various brain regions in MTLE patients were compared to healthy controls. The findings indicate that MTLE patients exhibit significant disruptions in functional connectivity, particularly involving the mesial temporal structures and associated networks. These disruptions were linked to impairments in cognitive functions such as memory and executive processes. However, most existing literature, including the studies mentioned above, focuses on correlating cognitive measures with static summaries of dynamic brain imaging or brain network data, often neglecting the dynamic pattern changes that occur over time. This static approach can overlook critical information about the temporal dynamics and fluctuations within brain networks that are essential for a comprehensive understanding of cognitive processes. The investigation of dynamic aspects of brain networks could provide deeper insights into the neural mechanisms underlying cognitive functions and their alterations in conditions like TLE.

\citet{neubauer.2009} explored the concept of neural efficiency, which posits that more intelligent individuals exhibit lower levels of brain activation during cognitive tasks compared to less intelligent individuals. This efficiency is particularly evident in tasks requiring executive functions, such as problem-solving and reasoning. The study reviewed evidence from various neuroimaging studies, including EEG and fMRI, demonstrating that higher intelligence is associated with more efficient neural processing, characterized by reduced activation in the prefrontal cortex and other brain areas involved in high-level cognitive functions. In this paper, we demonstrate that similar conclusions can be drawn from functional brain networks at rest obtained from rs-fMRI.  We will explore how these networks differ in TLE patients compared to normal controls and how these differences correlate with general intelligence. To address the limitations of existing methods, we developed a novel  topological embedding framework for time varying network data called {\em Topological Phase Diagram} (TPD) in aiding the topological modeling dynamic brain networks of temporal lobe epilepsy (TLE). The embedding is based on the Wasserstein distance between persistent diagrams, which provides the topological profile of data into 2D scatter plots. The Wasserstein distance or Kantorovich–Rubinstein metric, as  originally defined between probability distributions, is not scalable \citep{shi.2016,su.2015,ma.2023,chung.2023.NI,vallender.1974,canas.2012,berwald.2018}. We directly establish the relationship between the Wasserstein distance and edge weights in networks  making the method scalable and more adaptable. We achieve $\mathcal{O}(p \log p)$ run time in most graph manipulation tasks such as matching and averaging. Such scalable computation enables us to perform a computationally demanding task such as topological embedding and clustering with ease. We apply TPD  to elucidate the state space of dynamically changing functional brain networks, as captured through resting-state functional magnetic resonance imaging (rs-fMRI). The method allows us to delve into the temporal dynamics of brain connectivity, providing a nuanced understanding of how brain networks evolve over time.

\section{Methods}

%
%
%

\subsection{Preliminary: Graph filtrations}

We will represent a dynamically changing brain network represented as a weighted graph $\mathcal{X}(t) = (V, w(t))$, where $V=\{1, 2, \cdots, p\}$ is the set of nodes corresponding to different regions of the brain and $w(t) = (w_{ij}(t))$ represents the time-dependent edge weights between nodes $i$ and $j$. We may assume smaller $w_{ij}$ implies weaker connection while larger $w_{ij}$ implies stronger connection.

At each time point $t$, we can construct a binary network $\mathcal{X}_\epsilon(t) = (V, w_\epsilon(t))$ where the binary edge weights $w_{\epsilon,ij}(t)$ are given by:

\[ w_{\epsilon,ij}(t) =   \begin{cases} 1 &\; \mbox{  if } w_{ij}(t) > \epsilon;\\ 0 & \; \mbox{ otherwise}. \end{cases} \]

Here, $\epsilon$ acts as a threshold, determining the presence or absence of an edge based on the weight $w_{ij}(t)$ above given $\epsilon$. This process results in a sequence of nested graphs as the threshold $\epsilon$ is varied, encapsulating the multiscale structure. For a set of sorted edge weights 
$$w_{(1)}(t) < w_{(2)}(t) < \cdots < w_{(q)}(t)$$ 
at time $t$, the {\em graph filtration} is given by \citep{chung.2013.MICCAI,lee.2011.MICCAI,lee.2012.tmi}
\[ \mathcal{X}_{w_{(1)}}(t) \supset \mathcal{X}_{w_{(2)}}(t) \supset \cdots \supset \mathcal{X}_{w_{(q)}}(t). \]
We used the order statistic notation $w_{(i)}$ indicating the $i$-th smallest edge weight. For $p$ number of nodes in a complete graph, there are $q = (p^2-p)/2$ number of edges. This sequence of nested graphs, known as the graph filtration, captures the hierarchical organization of the network's connectivity. 

At the initial stage of the graph filtration includes a larger number of weaker connections. As we incrementally increase the threshold $\epsilon$, the graph begins to shed these weaker connections, progressively honing in on the more robust, significant interactions within the network. This sequence portrays a spectrum of networks ranging from most dense, representing the fully connected network, to the most sparse, which retains only the most significant connections. This approach allows for a nuanced understanding of the network's hierarchical structure, illustrating how connectivity patterns evolve as the threshold changes. By analyzing these filtrations over time, we can trace the dynamic evolution of brain connectivity, identifying which connections are persistent and robust under various conditions and which are transient.

In graph filtration, the 0D persistent homology tracks the evolution of connected components (0-cycles) across different thresholds. A connected component is born when a new edge creates a connection that is not part of an existing component. Once born, these components persist throughout the filtration, reflecting the non-decreasing nature of the 0th Betti number $\beta_0$ with increasing threshold values:

\[\beta_0( w_{(1)}) \leq \beta_0( w_{(2)})  \leq \cdots \leq \beta_0(w_{(q)}).\]

The 1D persistent homology, on the other hand, captures the formation and dissolution of loops or cycles. As the filtration progresses and more edges are deleted, some loops are destropyed, leading to their demise. This process is characterized by a non-increasing 1st Betti number $\beta_1$ \citep{chung.2019.ISBI}: 

\[\beta_1(w_{(q)}) \geq \cdots \geq \beta_1(w_{(1)}).\]

\begin{figure}[t]
\begin{center}
\includegraphics[width=1\linewidth]{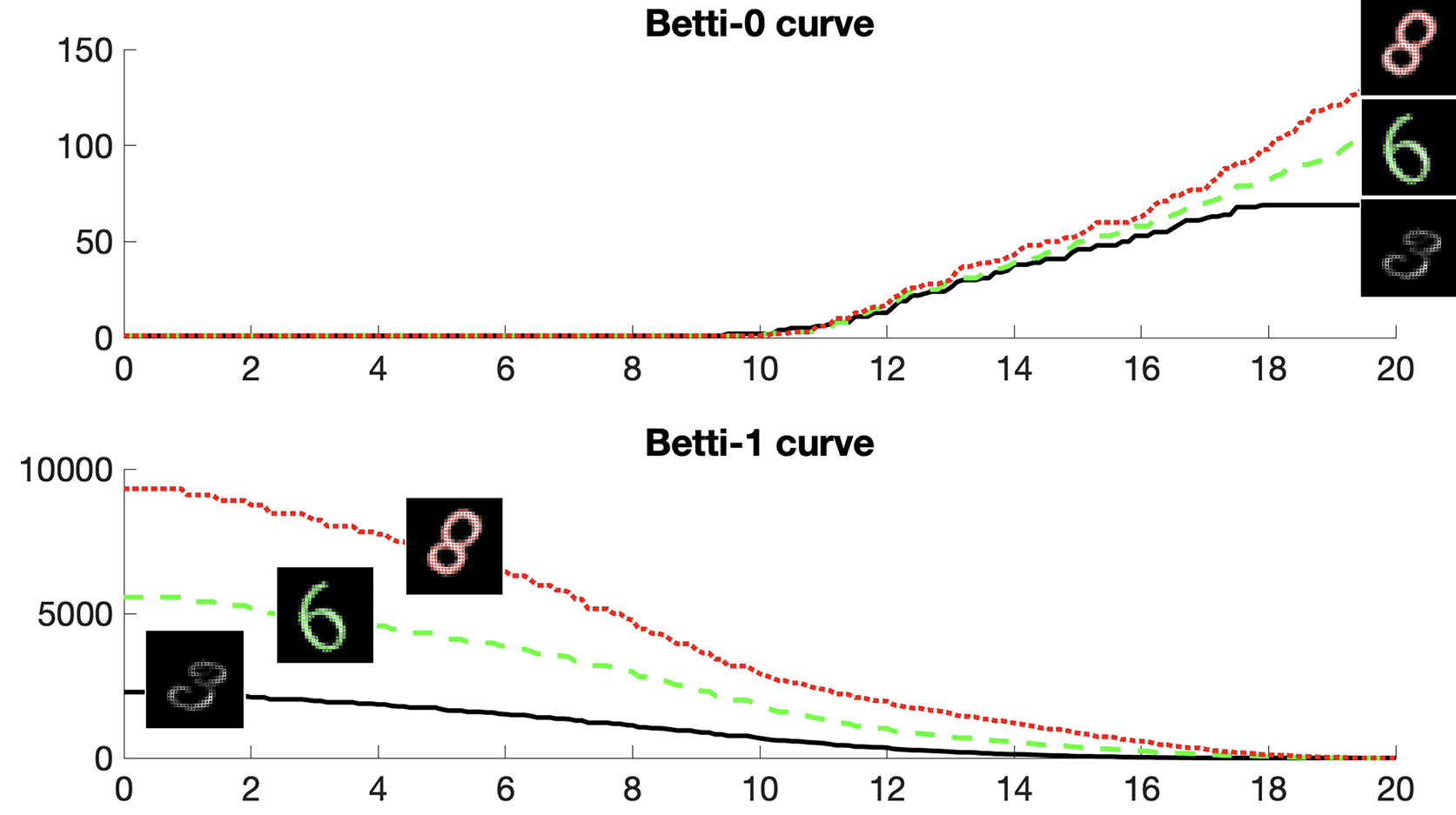}
\caption{Graph filtration is applied to digits in MINST  (Modified National Institute of Standards and Technology database) showing Betti number differences over the filtration. A graph representing each digit is used. Digits 3, 6 and 8 have 0, 1 and 2 cycles and thus they are topologically distinct.}
\label{fig:MINSTfiltration}
\end{center}
\end{figure}

%
%

\begin{figure}[t]
\begin{center}
\includegraphics[width=1\linewidth]{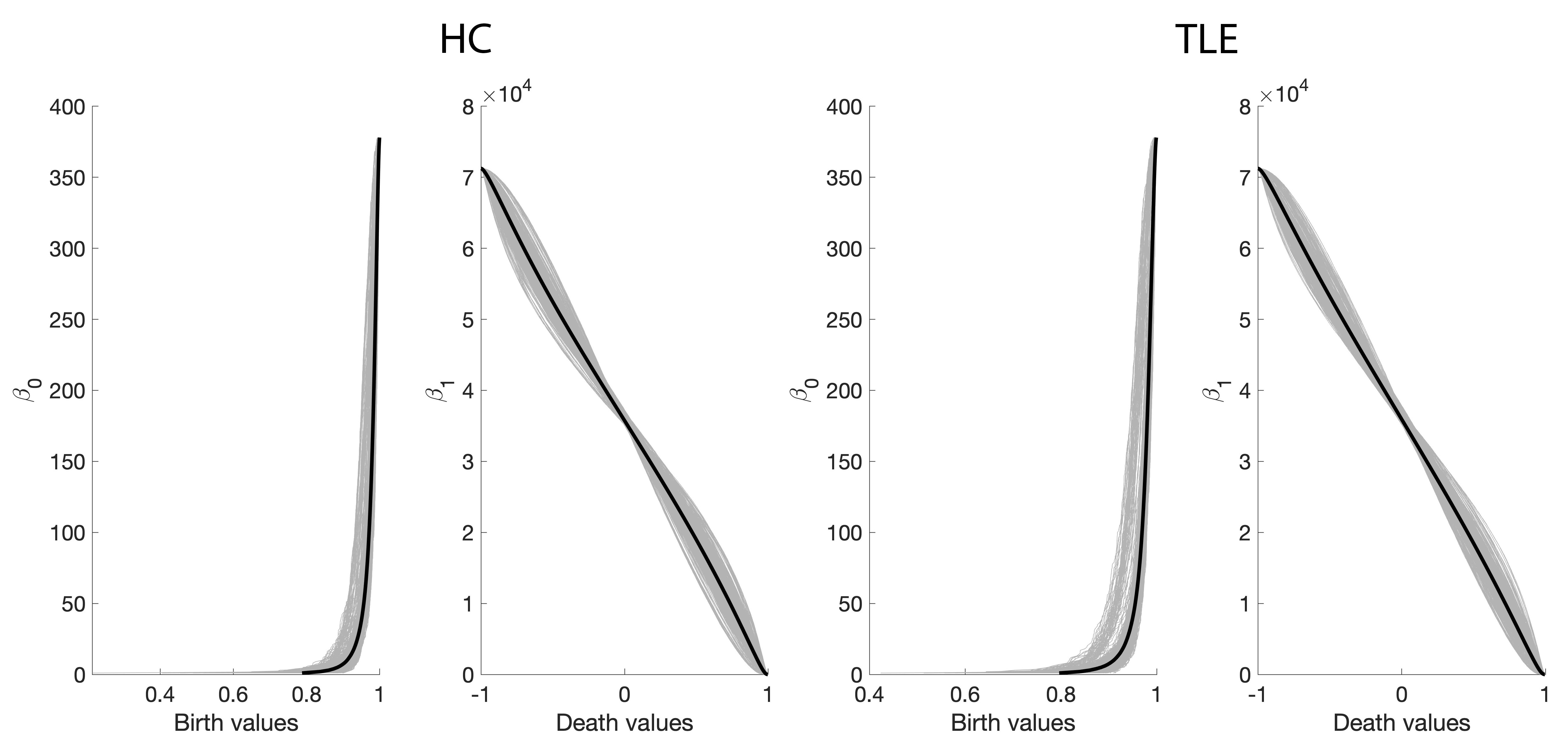}
\caption{Betti curves for one representative subject from HC and TLE obtained from the sliding window of size 20 TRs. For each group, Betti curves for each of the 1425 sliding windows are plotted in gray, while the average Betti curve across all sliding windows is depicted in black.}
\label{fig:betti_HCTLE}
\end{center}
\end{figure}

Figure \ref{fig:MINSTfiltration} displays an example of how $\beta_0$ and $\beta_1$ changes over filtration values for 3 topologically distinct MINST digits while Figure \ref{fig:betti_HCTLE} displays the Betti curves for two representative subjects in our study. In graph filtration, the persistence of 0-cycles is captured solely by their birth times, while cycles are uniquely identified by their death times. This unique aspect allows for a clear partition of the edge weight set $W$ into sets of birth values $W_b$ and death values $W_d$, which correspond to 0D and 1D topological features, respectively \citep{song.2021.MICCAI,song.2023}. The birth set $W_b$ is  related to the concept of a maximum spanning tree (MST) in a graph  \citep{lee.2012.tmi}. Each edge contributing to a 0-cycle in the MST signifies the emergence of a new connected component. The death set $W_d$, conversely, includes edges whose inclusion in the network closes a loop, signaling the termination of a cycle. When considering the temporal dynamics of a brain network, these birth and death sets can provide invaluable insights into how network connectivity evolves. For instance, changes in $W_b$ over time can indicate alterations in the network's integration or segregation, while variations in $W_d$ might reflect the network's resilience or vulnerability to perturbations.

\subsection{Topological Distances between Brain Networks}
\label{sec:dist}

\begin{figure}[t]
\begin{center}
\includegraphics[width=1\linewidth]{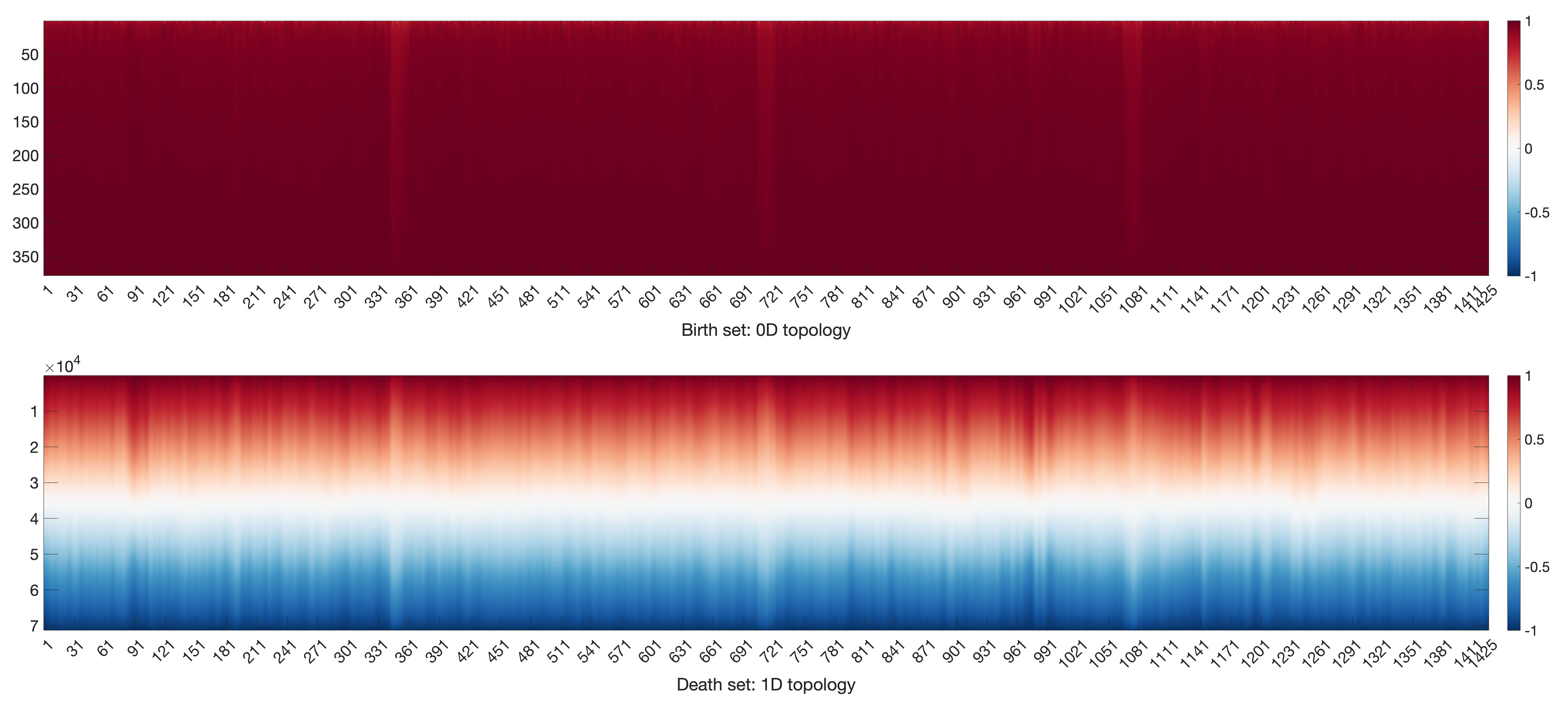}
\caption{Birth (top) and death (bottom) sets of the time varying correlation brain network for a representative subject. The horizontal axis represents time point.
The rs-fMRI brain networks are mainly characterized by 1D topology (cycles).}
\label{fig:BD-timeseries}
\end{center}
\end{figure}

 Persistent bars serve as a crucial tool for visualizing the evolution of topological features across varying filtration values \citep{ghrist.2008,topaz.2015}. These features, specifically connected components and loops, exhibit distinct behaviors in terms of their birth and death during the filtration process. In graph filtrations, we observe that 0D topological features, or connected components, emerge when an edge was disconnects two previously joint nodes at filtration value $b_{(i)}$. The set of these birth values forms the 0D persistence barcode $B = \{ b_{(i)} \}$ with each \(b_{(i)}\) indicating the birth of a new connected component (Figure \ref{fig:BD-timeseries}-top). Conversely, 1D topological features, or loops, are treated as having been born at \(-\infty\) in graph filtrations. A loop is considered to die when a deleted edge destroys the cycle. This death is captured at the filtration value equivalent to the weight of the edge causing the loop's demise. Accordingly, the 1D persistence barcode  consists of points \(D =  \{ d_{(i)} \} \), where each \(d_{(i)}\) marks the termination of a loop (Figure \ref{fig:BD-timeseries}-bottom).

During the graph filtration, we have  the unique decomposition of edge weights into sets \(W_b\) and \(W_d\), corresponding to the births of connected components and the deaths of loops, respectively. The emergence of a new component and the closure of a loop are mutually exclusive. With a complete graph encompassing \(p\) nodes, there are total  \(q = \frac{p(p-1)}{2}\) edges. Within this total, \(q_0 = p-1\) edges are associated with the creation of 0-cycles, equating to the number of edges in the graph's MST. Consequently, the remaining \(q_1 = \frac{(p-1)(p-2)}{2}\) edge weights are attributed to the eradication of 1-cycles. Thus, we have the following decomposition \citep{chung.2023.NI,song.2023}:

\begin{theorem}[Birth-death decomposition] For graph $\mathcal{X} = (V, w)$ with the edge weight set \( W  = \{ w_{(1)}, \cdots, w_{(q)} \} \) has the unique decomposition
\begin{equation}
\label{theorem:BDD}
W = B \cup D, \quad B \cap D = \emptyset
\end{equation}
where the birth set \( B = \{ b_{(1)}, b_{(2)}, \cdots, b_{(q_0)} \} \) is the collection of 0D sorted birth values, and the death set \( D = \{ d_{(1)}, d_{(2)}, \cdots, d_{(q_1)} \} \) is the collection of 1D sorted death values, with \( q_0 = p-1 \) and \( q_1 = \frac{(p-1)(p-2)}{2} \). Furthermore, the birth set \( B \) forms the 0D persistence  barcode, while the death set \( D \) forms the 1D persistence diagram.
\label{thm:decompose}
\end{theorem}

We  quantify and compare persistent barcodes across different instances or time points by leveraging the Wasserstein distance as a measure to compare persistent bars. Like the majority of statistical inference and learning  methods, where Euclidean distance is used  \citep{johnson.1967,hartigan.1979,lee.2012.tmi}, we propose to use topological distances. The Wasserstein distance, a probabilistic version of  optimal transport, provides a meaningful way to compare two probability distributions  
\citep{vallender.1974,canas.2012,berwald.2018}.

For distributions $X \sim f_1$ and $Y \sim f_2$, the $r$-Wasserstein distance $D_W$ is defined as:
\[ d (f_1, f_2) =  \left( \inf \mathbb{E} |  X - Y |^r \right)^{1/r}, \]
where the infimum spans all joint distributions of $X$ and $Y$ with marginals $f_1$ and $f_2$. This distance measures the minimum expected cost to transport mass from one distribution to another, offering a metric that satisfies key properties like positive definiteness, symmetry, and the triangle inequality.

Given two 0D persistent bars $B_1 = \{ b_{(i)}^1 \}$ and $B_2 = \{ b_{(i)}^2 \}$, we define their empirical distributions via Dirac delta functions:
\[ f_1 (x) = \frac{1}{q} \sum_{i=1}^q \delta (x-b_{(i)}^1), \quad f_2(x) = \frac{1}{q} \sum_{i=1}^q \delta (x-b_{(i)}^2). \]
The $2$-Wasserstein distance between $B_1$ and $B_2$ is then given by \citep{chung.2024.foundations}
\[ d_{0}(B_1, B_2)^2 =\sum_{i=1}^{q_0} (b_{(i)}^1 - b_{(i)}^2)^2.\]
and for 1D persistent bars, the $2$-Wasserstein distance $d_{1}$ is similarly given as
\[ d_{1}(D_1, D_2)^2 = \sum_{i=1}^{q_1} (d_{(i)}^1 - d_{(i)}^2), \]
where  $d_{(i)}^1, d_{(i)}^2$ are the ordered death values in persistent bars $D_1$ and $D_2$ respectively.  This computational efficiency stems from the fact that for graph filtrations, matching birth or death values align with the sorted order, simplifying the calculation. This allows the  distance computation {\em exactly} in $\mathcal{O}(q \log q)$  \citep{rabin.2011,song.2023,song.2021.MICCAI}.

To account for both 0D and 1D topological differences in brain networks, we use the sum of 0D and 1D Wasserstein distances $d^2 = d_0^2 + d_1^2$. Since it is unclear which feature contributes the most, equal weighting of 0D and 1D features ensures a balanced representation without bias towards either type of feature.

The passage describes a method for analyzing data through a topological phase diagram, focusing on the categorization of subjects into different health groups based on their topological features. Here's a reedited version of the paragraph with a focus on the topological phase diagram:

\subsection{Topological Phase Diagram}

\begin{figure}[t]
\begin{center}
\includegraphics[width=1\linewidth]{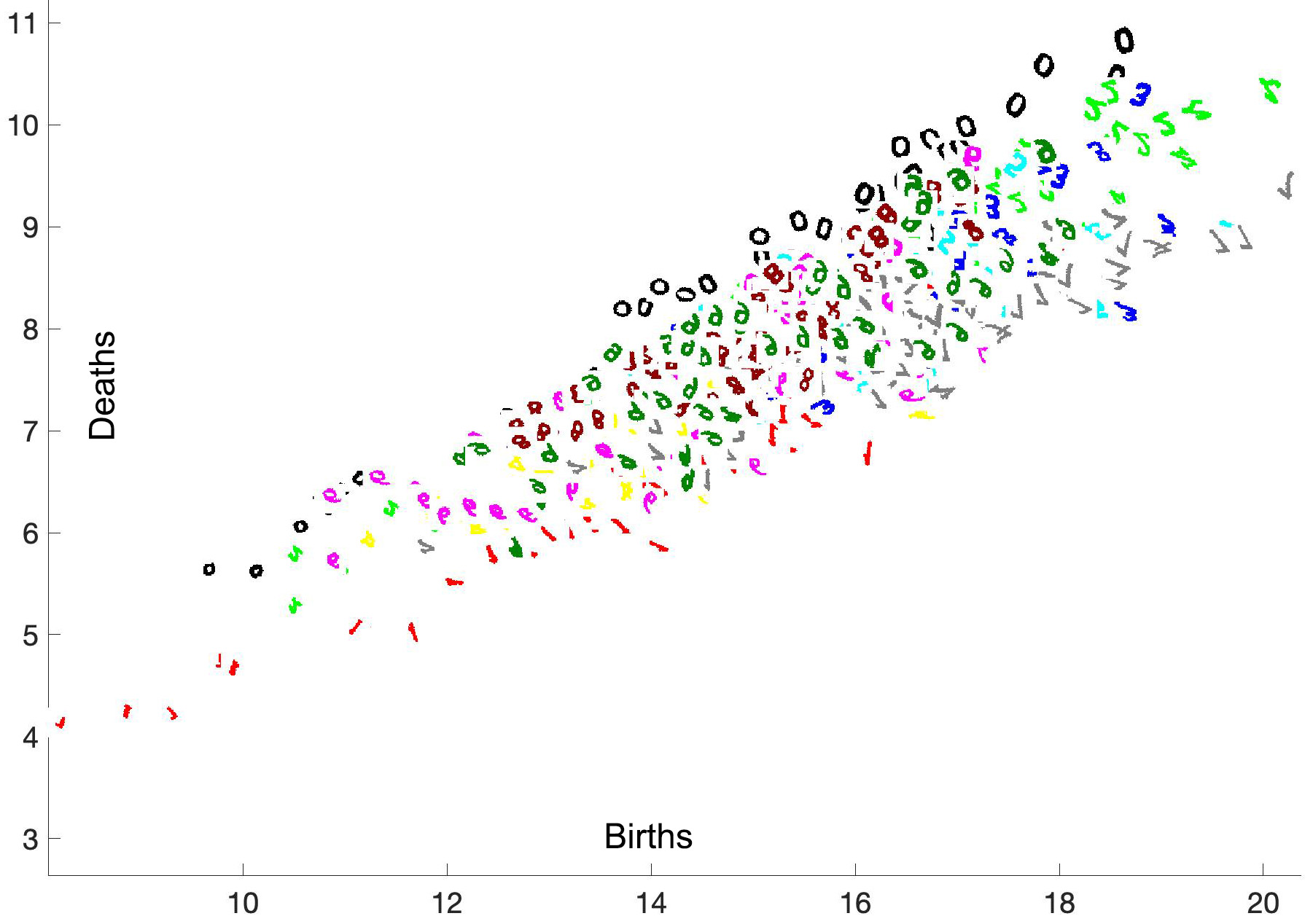}
\caption{Topological embedding of digits in MINST data on the Topological Phase Diagram (TPD). Topologically similar digits are clustered together along the diagonal direction. The embedding itself does not have any associated clustering cost. Still the method is able to cluster  them together.}
\label{fig:TPD-example}
\end{center}
\end{figure}

We introduce the {\em topological phase diagram} (TPD), a graphical representation that categorizes different states or phases of a system based on topological properties (Figures \ref{fig:TPD-example}). This diagram illustrates how the topological characteristics of a system or brain network evolve across different time points or clinical conditions. Using the birth-death decomposition, it is possible to decompose topology into 0D and 1D features. 
 In the topological phase diagram, the $x$-axis represents the spread with respect to the 0D topology, while the $y$-axis represents the spread with respect to the 1D topology \citep{chung.2023.NI}.

Consider the time series of brain networks with sorted birth and death values 
\[ b_{(1)}^t < b_{(2)}^t < \cdots < b_{(q_0)}^t \]
\[ d_{(1)}^t < d_{(2)}^t < \cdots < d_{(q_1)}^t \]
for time point $t$, where $t$ ranges from 1 to $T$. 
The embedding $x$- and $y$-coordinates for the brain network at time point $t$ are then given by  
\bqn x_t &=&    \frac{1}{q_0} \sum_{i=1}^{q_0} b_{(i)}^t.  \label{eq:x_t}\\
y_t &=&  \frac{1}{q_1} \sum_{i=1}^{q_1} d_{(i)}^t, \label{eq:y_t}
\eqn
which are the normalized {\em cumulative} birth an death values. Subsequently, the center of the embedding is given by
\bq \mu_b &=&  \frac{1}{T} \sum_{t=1}^T  \Big[ \frac{1}{q_0} \sum_{i=1}^{q_0} b_{(i)}^t \Big]\\
\mu_d &=&  \frac{1}{T} \sum_{t=1}^T  \Big[ \frac{1}{q_1} \sum_{i=1}^{q_1} d_{(i)}^t \Big]. 
\eq

\begin{figure}[t]
\begin{center}
\includegraphics[width=1\linewidth]{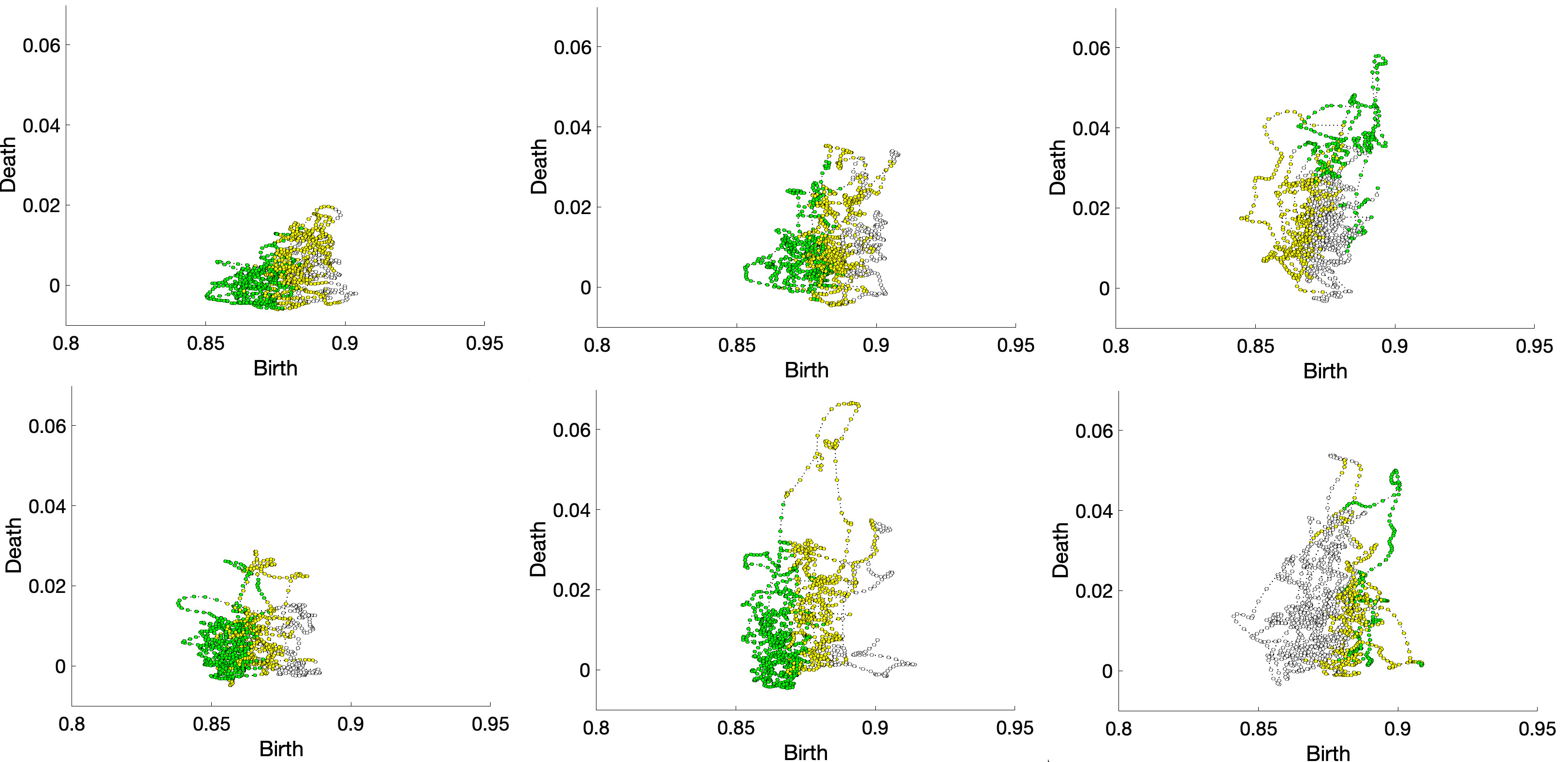}
\caption{Topological phase diagram (TPD) illustrating the temporal evolution of functional brain network topology for six representative subjects for the window size of 50 TRs (40 seconds). Topological clustering identifies transient states, each represented by a distinct color. We determined that $k=3$ is the optimal number of clusters to represent these states.}
\label{fig:TPD-coord2}
\end{center}
\end{figure}

\begin{figure}[t]
\begin{center}
\includegraphics[width=1\linewidth]{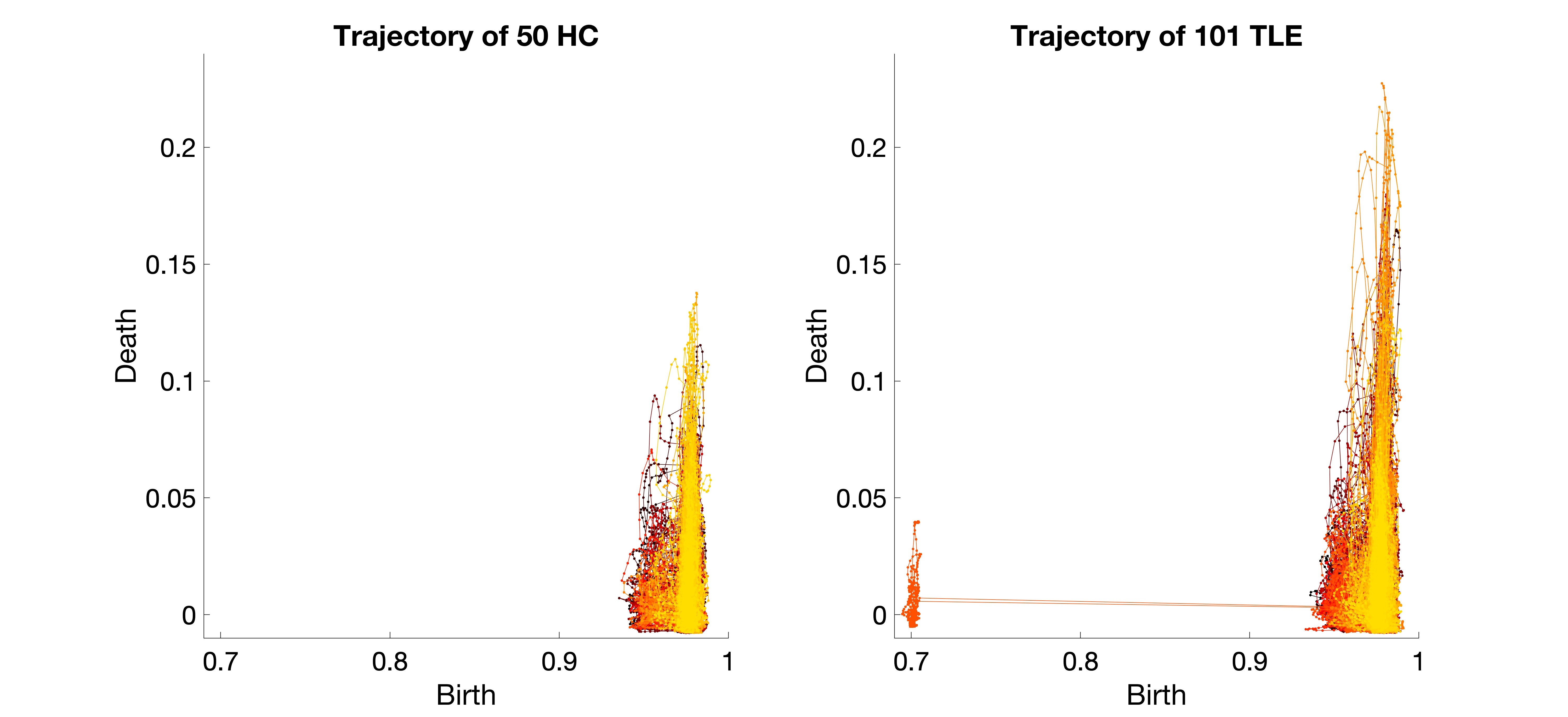}
\caption{Superimposition of TPD for a window size of 20 TRs (16 seconds) across all 50 HC and 101 TLE subjects, with each subject represented by a unique color. Consistent patterns are observable within each population. TLE patients exhibit a much wider range of death values, suggesting that to traverse longer trajectories, one must move more rapidly. Consequently, TLE is likely to exhibit more fluctuating connections and disconnections, forming cycles in 1D homology. One subject in the TLE group demonstrates definitively outlying behavior.}
\label{fig:TPD-curve}
\end{center}
\end{figure}

Figure \ref{fig:TPD-coord2} displays TPD for 6 representative subjects. If two groups of brain networks are analyzed, the  TPD should reveal a discernible difference in topological spread between the groups (Figure \ref{fig:TPD-curve}). In this study, we investigated if TPD can be effective used to characterize the time evolution of resting-state functional brain networks and use TPD to distinguish between different health conditions.

\begin{figure}[t]
\begin{center}
\includegraphics[width=1\linewidth]{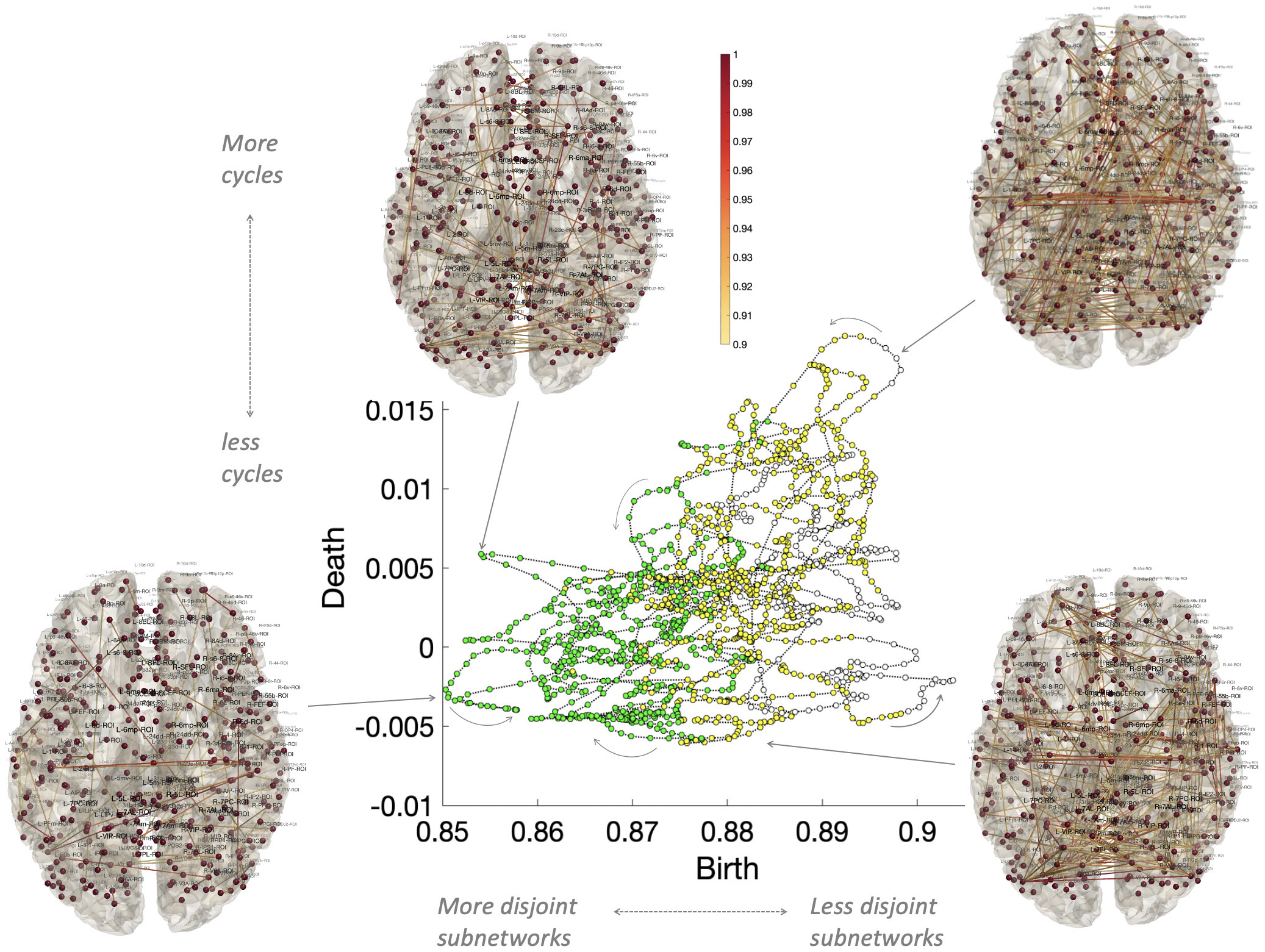}
\caption{Topological phase diagram (TPD) illustrating the temporal evolution of a functional brain network's topology for one subject over a window size of 50 TRs (40 seconds). Topological clustering reveals three transient states, each denoted by a unique color. Temporal progression is indicated with arrows. Overall, the network's evolution in the TPD predominantly follows a counterclockwise direction, although local, short-term deviations may occasionally proceed in a clockwise manner.}
\label{fig:TPD-example}
\end{center}
\end{figure}

\subsection{Random Field theory on Topological Phase Diagrams}

We developed a statitical inference procedure for directly analyzing a collection of TPD. Consider points $z_1, \cdots, z_n \in \mathbb{R}^2$ representing a topological phase diagram of brain network $\mathcal{X}(t)$. Their empirical distribution is given by
$$f(z) = \frac{1}{n}\sum_{i=1}^n \delta(z - z_i)$$
with the Dirac delta function $\delta(x)$. Note  
$\int_{\mathbb{R}^2}  f(z) \; dz = 1,$
ensuring $f$ acts as a probability density function. By applying Gaussian kernel smoothing with kernel  $K_{\sigma}(z) = \frac{1}{2\pi \sigma^2} \exp\left(-\frac{\|z\|^2}{2\sigma^2}\right),$
we obtain the kernel density estimation  \citep{fan.1996}:
$$K_{\sigma} * f(z) = \frac{1}{n}\sum_{i=1}^n K_{\sigma} (z - z_i).$$
This process is equivalent to solving the heat diffusion equation after time $ t= \sigma^2/2$:
$$\frac{\partial g(z,t)}{\partial t} = \Delta g(z,t),$$
with initial condition $g(z, t=0) = f(z)$ and Laplacian $\Delta$.

Given the smoothed TPD for each subject, we can analyze them collectively by computing the mean and standard deviation across subjects, which can reveal common patterns or variations across subjects. The smoothed TPD is used in performing a statistical inference  by testing the equivalence of TPD at each  birth and death value using the $t$-field $T(x)$ over square $S$, where points of TPD are defined \citep{chung.2020}. We are interested in determining the significance of TPD signals in square $S \subset \mathbb{R}^2$. 
 Since statistical test has to be done in every points in $S$, the multiple comparisons correction is needed. 
 For continuous functional data,  the random field theory  is often used \citep{andrade.2001,taylor.2007,worsley.1995,worsley.2004}. The random field theory assumes the measurements to be a smooth Gaussian random field. Kernel smoothing will make the data more smooth and Gaussian and increase the signal-to-noise ratio \citep{chung.2005.NI,lerch.2005.ni,wang.2010,yushkevich.2008}.

\begin{figure}[t]
\begin{center}
\includegraphics[width=1\linewidth]{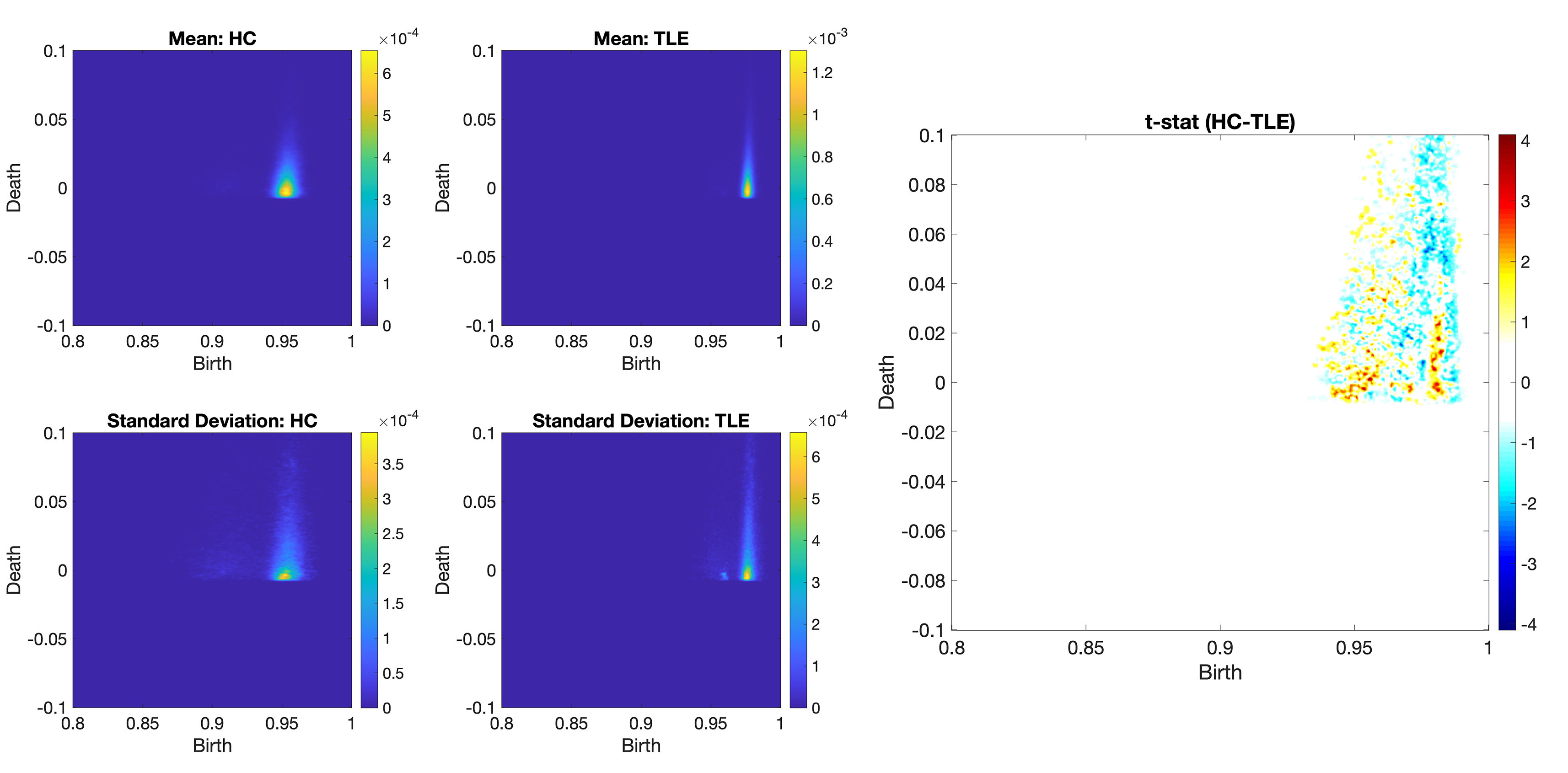}
\caption{Left: mean and standard deviation of smoothed TPD for window size 20 TRs = 16 seconds with kernel bandwidth 0.01. The consistency of the pattern across groups indicates the stability of the feature. The slight differences observed between the groups are likely to contribute to the statistical significance in group differences. Right: $t$-random field (HC - TLE). After the random field theory based multiple comparisons correction,  only  the red regions are statistically significant at 0.01 level. These regions corresponds to the states 1 and 2 during the state change. There is no signal detected in state 3.}
\label{fig:TPDstat1}
\includegraphics[width=1\linewidth]{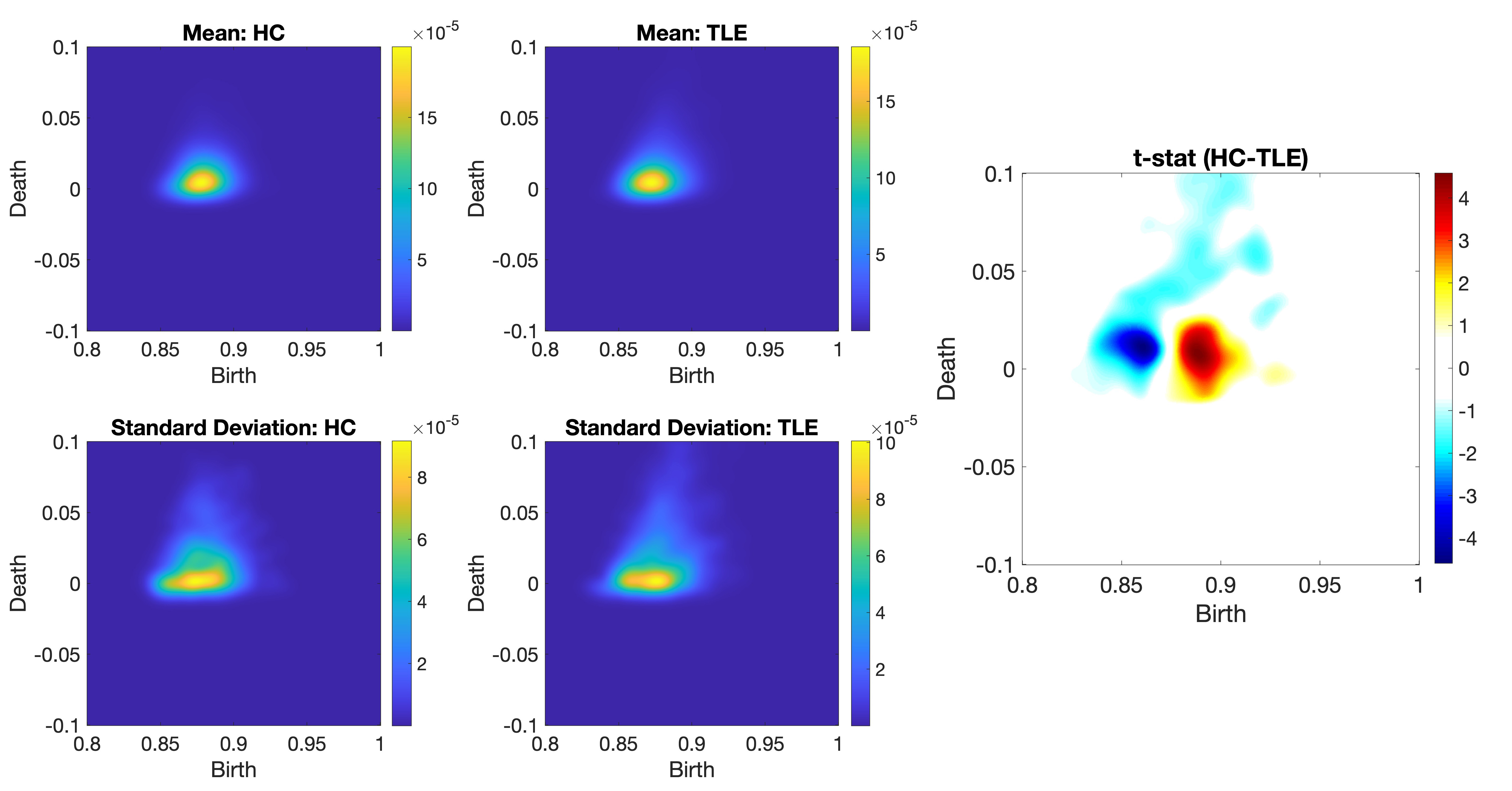}
\caption{Left: mean and standard deviation of smoothed TPD for window size 50 TRs = 40 seconds with kernel bandwidth 0.005. Right: $t$-random field (HC - TLE). Larger sliding window makes the resulting density maps far smother. After the random field theory correction,  Dark red and dark blue regions are statistically significant. These regions corresponds to the states 1 and 2 during the state change. There is no signal detected in state 3.}
\label{fig:TPDstat2}
\end{center}
\end{figure}

Consider functional measurements \( g_1, \ldots, g_n \) in square \( S \). The measurements can be modeled as 
\[ g_i (z) = h(z) + \epsilon_i(z), \]
where \( h \) is an unknown group-level signal, and \( \epsilon_i \) is a zero-mean Gaussian random field \cite{worsley.2004}. We are interested in determining the significance of \( h \), i.e.,
\begin{equation}
H_0: h(z) = 0 \mbox{ for all } z \in S \mbox{ vs. } H_1: h(z) > 0 \mbox{ for some } z \in S. \label{eq:hypothesis}
\end{equation}
Note that any point \( p_0 \) in \( S \) that gives \( h(p_0) > 0 \) is considered a signal. The hypothesis (\ref{eq:hypothesis}) is a multiple comparisons problem for continuously indexed hypotheses over the domain \( S \). Subsequently, a test statistic is given by a $t$-field \( T(z) \) for \( z \in S \). The resulting $t$-fields are given in Figures \ref{fig:TPDstat1} and \ref{fig:TPDstat2}.

The statistical significance after multiple comparisons correction is determined by the excursion probability \( P (\sup_{z \in S} T(z) \geq u) \), which is approximated by computing the expected Euler characteristic \( \chi \) of the random excursion set \( S_u = \{ x \in S : T(z) > u \} \) \citep{adler.1981, cao.2001, taylor.2007, worsley.2003}:
\[ P \Big(\sup_{x \in S} T(z) > u \Big) \doteq  \mathbb{E} \chi (S_u) = \sum_{j=0}^2 \mu_j(S) \rho_j(u), \]
where \( \mu_j(S) \) is the \( j \)-th Minkowski functional (also known as Lipschitz-Killing curvatures or intrinsic volume) of \( S \), and \( \rho_j \) is the \( j \)-th Euler characteristic (EC) density of the T-field. The Minkowski functionals for square $S$ with side length $a$:
$$\mu_0(S) = 1, \quad \mu_1(S) = 2a, \quad \mu_2(S) = a^2.$$
These correspond to the Euler characteristic, the half the boundary length, and the area of the square, respectively.

The EC densities \( \rho_j(u) \) for the T-field with \( \nu \) degrees of freedom are given by:
\bq
\rho_0 (u) &=& 1 - P(T_{\nu} \leq u), \\
\rho_1 (u) & =& \frac{1}{\sqrt{2\sigma^2}} \cdot \frac{1}{2\pi} \left( 1 + \frac{u^2}{\nu} \right)^{-(\nu-1)/2}, \\
\rho_2 (u) & =& \frac{1}{2\sigma^2} \cdot \frac{1}{(2\pi)^{3/2}} \frac{\Gamma \left(\frac{\nu +1}{2} \right)}
{\left( \frac{\nu}{2} \right)^{1/2} \Gamma \left(\frac{\nu}{2} \right)} u \left( 1 + \frac{u^2}{\nu} \right)^{-(\nu -1)/2}. 
\eq
The EC-density has the kernel bandwidth $\sigma$ in the formulation so the inference is done at a particular smoothing scale \citep{worsley.2004,adler.1981,taylor.2007}. 
The expected Euler characteristic is coded in MATLAB as {\tt EC\_t\_square.m}, which inputs the side of square $a$, bandwidth $\sigma$ and threshold $u$.

\subsection{Topological Clustering of  State Space }
\label{sec:clustering}

We introduce a novel topological approach for estimating state spaces in dynamically changing functional human brain networks derived from rs-fMRI. Our approach, distinct from traditional methods that consider collections of graphs, deals with a time series of graphs \(\mathcal{X}(t)\), capturing the evolving connectivity over time. We employ the Wasserstein distance to cluster these time-varying brain networks into topologically distinct states \citep{mi.2018,yang.2020}, incorporating the temporal dimension of the data. Specifically, we aim to cluster the time series of graphs \(\mathcal{X}(1), \mathcal{X}(2), \ldots \) into \(k\) clusters \(C = (C_1, \ldots, C_k)\) such that
$$\bigcup_{i=1}^k C_i = \{\mathcal{X}(1), \mathcal{X}(2), \ldots \}, \quad C_i \cap C_j = \emptyset \text{ for } i \neq j.$$

Let \(\mu_j\) denote the topological cluster centroid within \(C_j\), defined as
\bqn \mu_j = \arg \min_{Y \in C_j} \sum_{\mathcal{X}(t) \in C_j} d^2(Y, \mathcal{X}(t)),
\label{eq:centroid}
\eqn
where \(d^2 = d_0^2 + d_1^2\) represents the combined Wasserstein distance. The within-cluster distance from the cluster centroid $\mu = (\mu_1, \cdots, \mu_k)$ is
$$l_W (C; \mu) = \sum_{j=1}^k \sum_{\mathcal{X}(t) \in C_j} d^2( \mathcal{X}(t), \mu_j).$$
The optimal cluster is found by minimizing within-cluster distance $l_W(C;\mu)$  over every possible partition of $C$. The proof for the local convergence is given in  \citet{chung.2023.NI}.

Similar to the $k$-means clustering, which only guarantees convergence to a local minimum\citep{huang.2020.NM}, our topological clustering algorithm does not ensure convergence to the global minimum. The choice of initial cluster centroids can significantly influence the results, as the algorithm might get trapped in a local minimum and fail to reach the global minimum. To address this issue, we repeat the algorithm multiple times with different random seeds, ultimately selecting the run with the smallest minimum. The method is implemented in the MATLAB function {\tt WS\_cluster.m}, which takes a series of networks as input and outputs the corresponding cluster labels and clustering accuracy.

\subsubsection{Topological centroid}

For dynamically changing functional brain networks represented as a time series of graphs \(\mathcal{X}(t)\), we introduce an approach to identify the topological mean within each cluster of these networks. This concept is akin to the Wasserstein barycenter \citep{agueh.2011,cuturi.2014} and the Fréchet mean \citep{le.2000,turner.2014,zemel.2019,dubey.2019}. Unlike conventional averaging methods that focus on element-wise arithmetic means of edge weights, our method leverages the Wasserstein distance to capture the underlying topological features of the networks. However, this approach, while capturing the average connectivity strength, may not accurately reflect the central topological tendencies \citep{chung.2023.NI}.

Given a cluster \(C_j\) comprising a time series of brain networks \( \mathcal{X}(t_{j1}), \ldots, \mathcal{X}(t_{jm})\), the topological centroid \(\mu_j\) is defined as the network that minimizes the sum of squared Wasserstein distances to all other networks within the cluster in (\ref{eq:centroid}). This can be approximated by relaxing the condition such that the centroid is not constrained within the cluster:
\bqn \mu_j = \arg \min_{Y} \sum_{k=1}^m d^2(Y, \mathcal{X}(t_{j_k}))
\label{eq:centroid_relaxed}, \eqn
which can be solved exactly as follows.

\begin{figure}[t]
\begin{center}
\includegraphics[width=0.7\linewidth]{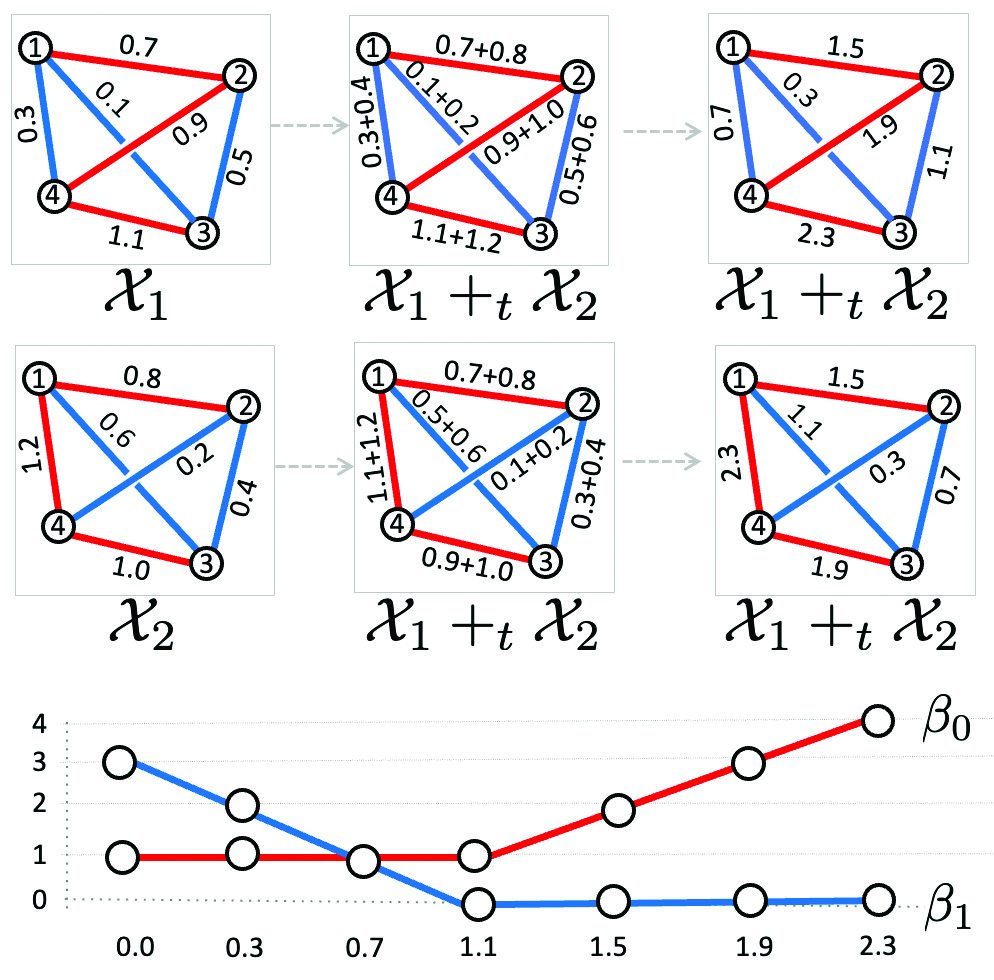}
\caption{Topological addition $+_t$ on two graphs $\mathcal{X}_1$ and $\mathcal{X}_2$. The resulting graph $\mathcal{X}_1 +_t \mathcal{X}_2$ is constructed by summing the matched sorted edge weights in MST (birth set colored red) and none-MST (death set colored blue) separately. $\mathcal{X}_1 +_t \mathcal{X}_2$ is not unique as shown in this example. The resulting birth and death values are projected onto the MST and none-MST edges of the original graphs. The resulting graphs (top right and bottom right) yield the identical $\beta_0$ and $\beta_1$ curves, thus the topological distance cannot discriminate them and should be treated as topologically equivalent.}
\label{fig:addition}
\end{center}
\end{figure}

\begin{figure}[t]
\begin{center}
\includegraphics[width=1\linewidth]{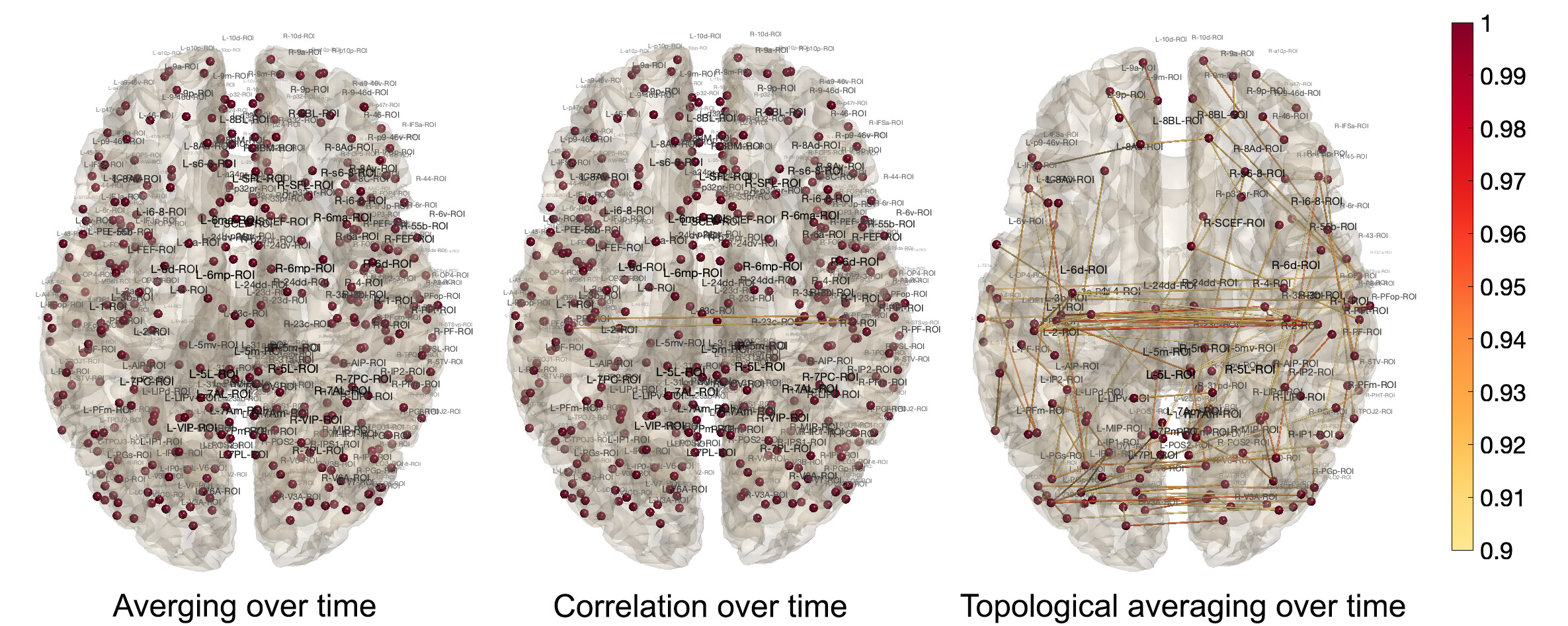}
\caption{Left: Connectivity matrix averaged across all time points for the subject shown in Figure \ref{fig:TPD-coord1}. Averaging rs-fMRI correlation matrices over the entire duration can dilute the signal. Middle: The conventional approach of computing the Pearson correlation across all time points to derive a static summary. Some highly connected connections at specific times simply disapper. Right: Topological averaging of correlation matrices across all time points. By employing birth-death decomposition for averaging, we preserve more underlying connectivity pattern.}
\label{fig:averging}
\end{center}
\end{figure}

Consider graphs \(\mathcal{X}_i = (V, w^i)\) with the corresponding birth-death decomposition $W_i = B_i \cup D_i$. Let $B_i = \{ b_{(1)}^i, \cdots, b_{(q_0)}^i \}$ and $D_i = \{ d_{(1)}^i, \cdots, d_{(q_1)}^i \}$ be sorted birth and death sets for the $i$-th graph. We then define the \textit{topological addition} to amalgamate two graphs, \(\mathcal{X}_1 = (V, w^1)\) and \(\mathcal{X}_2 = (V, w^2)\), into \(\mathcal{X}_1 +_t \mathcal{X}_2 = (V, w)\). The  birth-death decomposition of $\mathcal{X}_1 +_t \mathcal{X}_2$ is given by
$$B \cup D = (B_{1} + B_{2}) \cup (D_{1} + W_{2}),$$
where $$B_1 + B_2 = \{ b_{(1)}^1 + b_{(1)}^2, \cdots, b_{(q_0)}^1 + b_{(q_0)}^2 \}$$ and $$D_1 + D_2 = \{ d_{(1)}^1 + d_{(1)}^2, \cdots, d_{(q_1)}^1 + d_{(q_1)}^2\}$$ are element-wise addition of sorted order sets. To realize and visualize such a graph, we need to determine the corresponding edge weight matrix \( w \). However, such graphs are not unique. For graphs with \( p \) labeled nodes, the number of distinct graphs yielding the same birth-and-death decomposition corresponds to the total number of spanning trees. Cayley's formula states that the total number of spanning trees for a complete graph with \( p \) nodes is \( p^{(p-2)} \) \citep{cayley.1878}. Figure \ref{fig:addition} illustrate the topological addition of two graphs, where the added birth and death values are simply projected onto the original graphs.  Subsequently, we define {\em the topological mean} as \( (\mathcal{X}_1 +_t \mathcal{X}_2)/2 \), where the corresponding birth-death set is scaled by $1/2$. Then we can prove that the cluster mean $\mu_j$ is given by the topological mean:
\begin{theorem}
\label{theorem:sum}
The topological mean $$\mu_j = \frac{\mathcal{X}(t_{j1}) +_t \ldots +_t \mathcal{X}(t_{jm})}{m}$$ of all the graphs in the cluster $C_j$
is the minimizer of 
$$\min_{Y} \sum_{k=1}^m d^2(Y, \mathcal{X}(t_{j_k})).$$
\end{theorem}
The proof is given in \citep{chung.2023.NI}. The  topological mean of graphs is the minimizer with respect to the topological distance, which is analogous to the sample mean as the minimizer of Euclidean distance. However, the topological mean of graphs is not unique in geometric sense. It is only unique in topological sense.  It is possible that many different graphs have the identical birth-death decomposition, and thus it may not possible to distinguish them topologically (Figure \ref{fig:addition}). Thus, we need to establish a sort of anchor graph onto which we can project the topological mean graph. This is implemented in the function {\tt WS\_project.m}, which takes the birth and death sets as inputs and projects them onto a given connectivity matrix. The resulting connectivity matrix will maintain the same topological structure given in the birth and death sets. Figure \ref{fig:averging} demonstrates the performance of topological averaging (right) against other existing approach for obtaining a static summary network. In terms of signal retention, topological averaging far outperforms existing methods. For topological averaging, we projected it to the Pearson correlation over the whole time point.

\subsubsection{Assigning consistent cluster labels across subjects}
Let \(y_t\) be the true cluster label for the brain network $\mathcal{X}(t)$ at the \(t\)-th time point. Let \(\widehat{y}_t\) be the estimate of \(y_t\) obtained from topological clustering. Define \(y=(y_1, \cdots, y_T)\) and \(\widehat{y}=(\widehat{y}_1, \cdots, \widehat{y}_T)\). There is no direct association between the true clustering labels \(y\) and the predicted cluster labels \(\widehat{y}\), as they are independent. For \(k\) clusters \(C_1, \cdots, C_k\), its permutation \(\pi(C_1), \cdots, \pi(C_k)\) is also a valid clustering for \(\pi \in \mathbb{S}_k\), the permutation group of order \(k\). There are \(k!\) possible permutations in \(\mathbb{S}_k\) \citep{chung.2019.CNI}. Therefore, it is crucial to obtain consistent cluster labels across subjects for any subsequent statistical analysis across subjects. 
Once we identify networks in cluster $C_j$, we can compute the  embedding coordinates $(\mu_b^j, \mu_d^j)$ of the cluster centroid of in TPD. In persistent homology, signals with longer persistence are considered as having a more significant fleeting topological signal. The quantity $\mu_d^j - \mu_b^j$ can be interpreted as a proxy for the overall persistence of the cluster. Consequently, we relabel the cluster with the largest value of $\mu_d^j - \mu_b^j$.  the lowest cluster label, ensuring consistency in the labeling across time points and subjects.

\subsection{Hodge Decomposition in Topological Phase Diagram}

We explore the dynamics of topological changes over time within a Topological Phase Diagram (TPD). Given points $(x_1, y_1), \cdots, (x_T, y_T)$ in TPD, we define a displacement vector \(w_k\) at each time point \(k\), representing the topological change between consecutive time points:
\[ w_k = (x_{k+1} - x_k, y_{k+1} - y_k).\] 
Then we obtain the vector field $w= \{ w_1, w_2, \cdots, w_{T-1} \}$, which serves as a proxy for the underlying dynamical processes driving the topological evolution of the brain network. We can study the properties of the vector field, such as its divergence and curl, which provide insights into the behavior of the topological changes, such as expansion, contraction, rotation, and shearing. For this, we obtain the smooth representation of the vector field $v(z)$ as
$$\frac{\partial v(z,t)}{\partial t}  = \Delta v (z,t)$$
with initial condition $v(z, t=0) = w$. Subsequently, we obtain smooth vector field $v$ which provides a smooth vector description of the topological changes over time.

Given a smooth vector field \(v(z)\) within a square domain \(S\), the Hodge decomposition enables the unique decomposition of the field into three orthogonal components: an irrotational (gradient) part, a solenoidal (curl) part, and a harmonic part, expressed as:
\[
v(z) = \nabla \phi(z) + \nabla \times \psi(z) + h(z),
\]
where \(\nabla \phi(z)\) represents the gradient of a scalar potential \(\phi(z)\), capturing the irrotational component of \(v(z)\), while \(\nabla \times \psi(z)\) represents the curl of a scalar field \(\psi(z)\), capturing the solenoidal or divergence-free component of \(v(z)\). This component highlights the areas where the vector field exhibits rotational behavior, independent of any outward or inward flow as characterized by the gradient component. The harmonic part, \(h(z)\), is both divergence-free and curl-free, embodying the component of \(v(z)\) that cannot be expressed as either a gradient or a curl. This part is particularly significant in the context of closed domains where boundary conditions play a role, and it captures the intrinsic properties of the vector field that are maintained across the domain.

\begin{figure}[t]
\begin{center}
\includegraphics[width=1\linewidth]{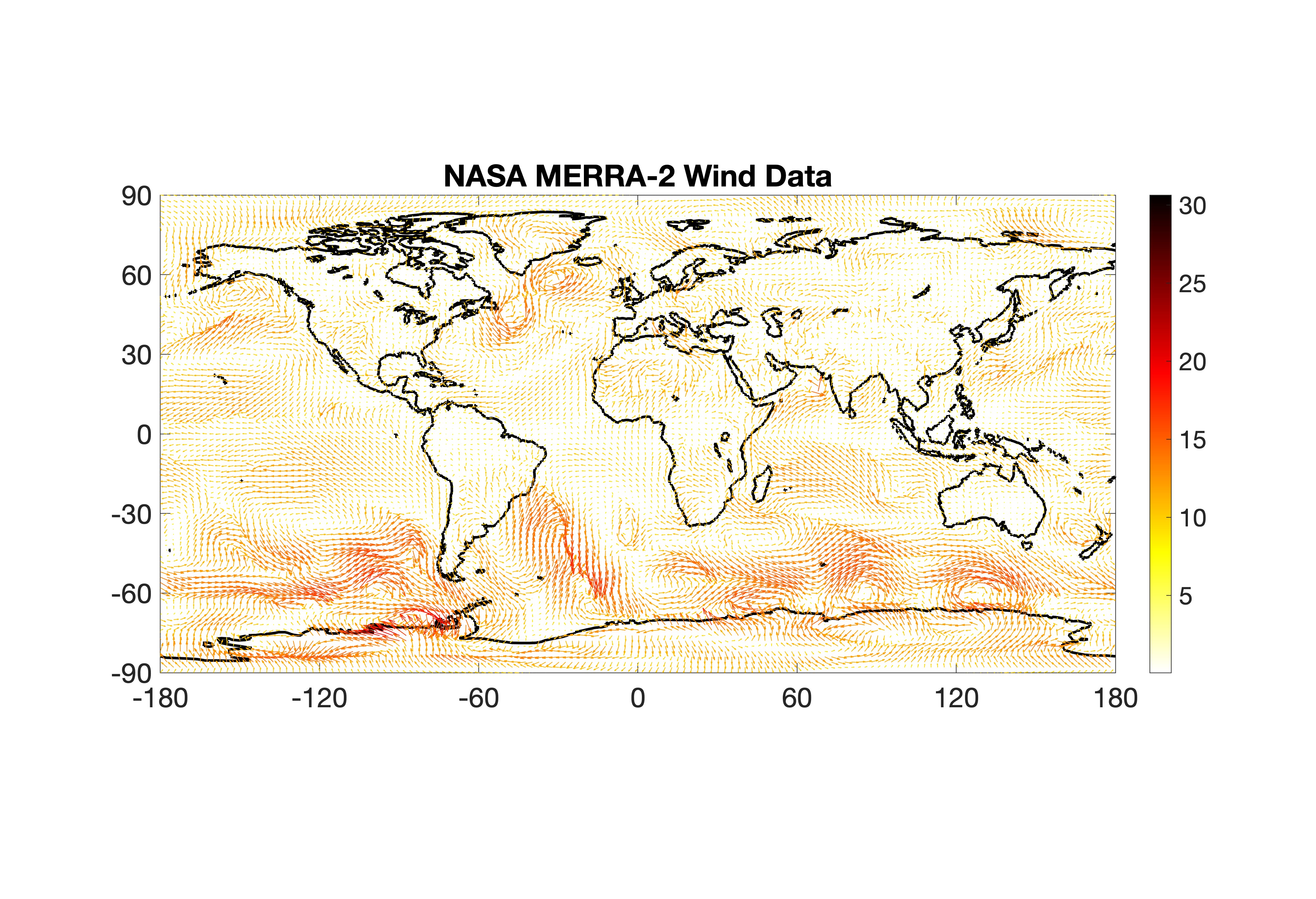}
\caption{NASA MERRA-2 wind data analyzed using the Hodge decomposition.}
\label{fig:NASA}
\end{center}
\end{figure}

We illustrate our methodology using an intuitive example from  NASA MERRA-2 dataset, a comprehensive collection of atmospheric data that offers a detailed  weather and climate data \citep{carvalho.2019}. We utilized the wind vector data in the dataset as an illustration. 

\subsubsection{Estimation of gradient field}
The gradient filed  \(\nabla \phi(z)\) in the coordinates \(z = (x, y)\) is given by 
\[
\nabla \phi(z) = \left( \frac{\partial \phi}{\partial x}, \frac{\partial \phi}{\partial y} \right).
\]
The potential $\phi$ can be estimated by minimizing the sum of squared differences between \(\nabla \phi(z)\) and \(v(z)\):
\[
\widehat{\phi} = \min_\phi \sum_{x \in S} \| \nabla \phi(z) - v(z) \|^2.
\]
Discretize the domain \(S\) into a finite set of points \(s_1, s_2, \cdots, s_n\). 
The vectorized form of the scalar potential \(\phi\) at these points is:

\[
\boldsymbol{\phi} = 
\begin{bmatrix}
\phi(s_1) \\
\phi(s_2) \\
\vdots \\
\phi(s_n)
\end{bmatrix}.
\]

\begin{figure}[t]
\begin{center}
\includegraphics[width=1\linewidth]{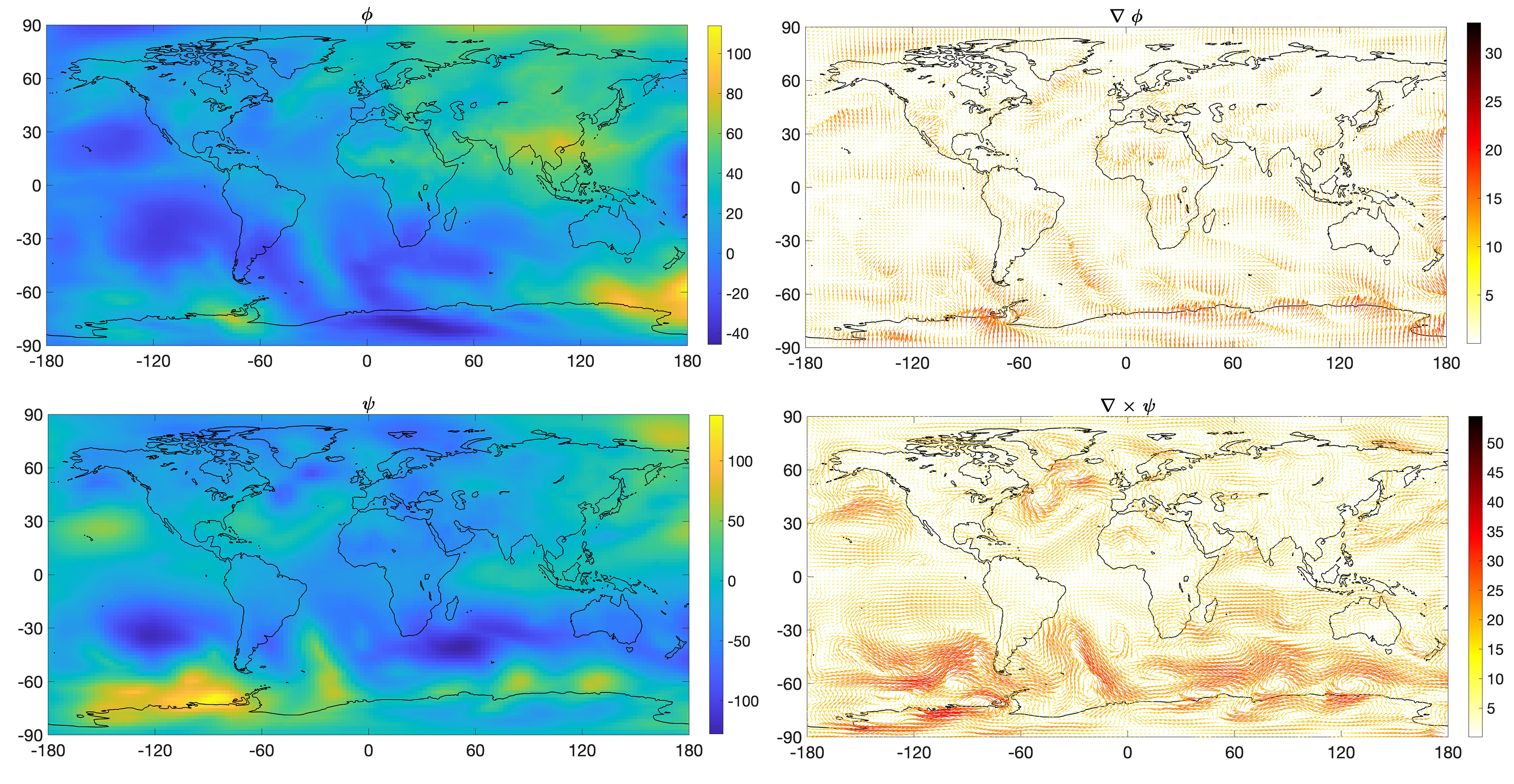}
\caption{NASA MERRA-2 wind data analyzed using the Hodge decomposition. Left: the estimated scalar field $\phi$ and $\psi$. The resulting gradient field $\nabla \phi$ and curl field $\nabla \times \psi$.}
\label{fig:NASA-vector}
\end{center}
\end{figure}

Representing the gradient operator $\nabla$ in a discretized form involves matrices \(\boldsymbol{\nabla}_x\) and \(\boldsymbol{\nabla}_y\) that approximate the partial derivatives with respect to \(x\) and \(y\), respectively, over grid. The vectorized form of \(v(x)\) is separated into components:
\[
\mathbf{v}_x = 
\begin{bmatrix}
v_x(s_1) \\
v_x(s_2) \\
\vdots \\
v_x(s_n)
\end{bmatrix},
\quad
\mathbf{v}_y = 
\begin{bmatrix}
v_y(s_1) \\
v_y(s_2) \\
\vdots \\
v_y(s_n)
\end{bmatrix}.
\]
Then we solve the simultaneous matrix equations 
\[
 \boldsymbol{\nabla}_x    \boldsymbol{\phi} = {\bf v}_x, \quad \boldsymbol{\nabla}_y \boldsymbol{\phi} = {\bf v}_y,
\]
which is equivalent to solving
$$
\begin{bmatrix}
\boldsymbol{\nabla}_x\\
\boldsymbol{\nabla}_y
\end{bmatrix} \boldsymbol{\phi} =  
\begin{bmatrix}  {\bf v}_x\\ 
 {\bf v}_y
\end{bmatrix}.
$$
Subsequently, we obtain the least squares estimation
\[
\widehat{\boldsymbol{\phi}} =  \Big( \begin{bmatrix}
\boldsymbol{\nabla}_x\\
\boldsymbol{\nabla}_y
\end{bmatrix}^{\top}   \begin{bmatrix}
\boldsymbol{\nabla}_x\\
\boldsymbol{\nabla}_y
\end{bmatrix} \Big)^{-1} \begin{bmatrix}
\boldsymbol{\nabla}_x\\
\boldsymbol{\nabla}_y
\end{bmatrix}^{\top} 
\begin{bmatrix}  {\bf v}_x\\ 
 {\bf v}_y
\end{bmatrix}.
\]
Once \(\widehat{\boldsymbol{\phi}}\) is computed, the gradient field \(\nabla \phi(x)\) can be further computed, providing an estimate of the irrotational component of \(v(x)\) (Figure \ref{fig:NASA-vector}).

\subsubsection{Estimation of curl and harmonic components}

For the curl component \(\nabla \times \psi(z)\), which captures the solenoidal or divergence-free aspect of \(v(z)\), we find the scalar potential \(\psi(z)\) such that \(\nabla \times \psi(z)\) optimally fits the solenoidal part of \(v(z)\). Let:

\[
\boldsymbol{\psi} = 
\begin{bmatrix}
\psi(s_1) \\
\psi(s_2) \\
\vdots \\
\psi(s_n)
\end{bmatrix}.
\]
The operation \(\nabla \times \psi(z)\) in 2D is defined as
\[
\nabla \times \psi(z) = \left( \frac{\partial \psi}{\partial y}, -\frac{\partial \psi}{\partial x} \right),
\]
Then we solve for the matrix equation
$$
\begin{bmatrix}
\boldsymbol{\nabla}_y\\
-\boldsymbol{\nabla}_x
\end{bmatrix} \boldsymbol{\psi} =  
\begin{bmatrix}  {\bf v}_x\\ 
 {\bf v}_y
\end{bmatrix}.
$$
The least squares estimation is then given by
\[
\widehat{\boldsymbol{\psi}} = \left( \begin{bmatrix}
-\boldsymbol{\nabla}_y \\
\boldsymbol{\nabla}_x
\end{bmatrix}^{\top} \begin{bmatrix}
-\boldsymbol{\nabla}_y \\
\boldsymbol{\nabla}_x
\end{bmatrix} \right)^{-1} \begin{bmatrix}
-\boldsymbol{\nabla}_y \\
\boldsymbol{\nabla}_x
\end{bmatrix}^{\top} 
\begin{bmatrix}
\mathbf{v}_x \\
\mathbf{v}_y
\end{bmatrix}.
\]
Once we estimate \(\widehat{\boldsymbol{\psi}}\), the estimated curl component 
is simply given as \(\nabla \times \widehat{\psi}(z)\) is computed using:

\[
\nabla \times \widehat{\psi}(z) = \begin{bmatrix}
\boldsymbol{\nabla}_y \widehat{\boldsymbol{\psi}} \\
-\boldsymbol{\nabla}_x \widehat{\boldsymbol{\psi}}
\end{bmatrix}.
\]
This vector field represents the solenoidal component of \(v(x)\), capturing the inherent rotational features in the original vector field (Figure \ref{fig:NASA-vector}).

\(h(x)\) is the harmonic component, which is both divergence- and curl-free. In a simply connected domain, \(h(x)\) is typically zero unless there are boundary conditions or other constraints. When $h(x)$ is not zero, it is obtained as residual 
$$\widehat h(x) = v(x) - \nabla \widehat \phi(x) - \nabla \times \widehat \psi(x).$$

\section{Results}

\subsection{Data and Preprocessing}

The study included 101 patients with temporal lobe epilepsy (TLE) and 50 healthy controls, all part of Epilepsy Connectome Project (ECP) database, a collaborative project between the Medical College of Wisconsin and the University of Wisconsin-Madison \citep{hwang.2020}. We used 50 healthy controls   (mean age 31.78 $\pm$ 10.32 years) and  101 chronic temporal lobe epilepsy (TLE) patients (mean age 40.23 $\pm$ 11.85).  TLE participants were with a Full-Scale Intelligence Quotient (FSIQ) above 70, fluent in English, and without medical contraindications to MRI \citep{garcia.2022}. The TLE diagnosis and seizure onset side were confirmed by a board-certified neurologist, following the International League Against Epilepsy criteria. Exclusion criteria were lesions other than mesial temporal sclerosis and any active infectious/autoimmune/inflammatory seizure etiology. Control participants were excluded if they had an Edinburgh Laterality Quotient under 50, were non-native English speakers, had a history of learning disabilities, brain injuries, substance abuse, major psychiatric illnesses, were taking vasoactive drugs, or had contraindications to MRI \citep{garcia.2022}. All participants provided written informed consent, with the study approved by the Medical College of Wisconsin's and University.

The resting-state fMRI were collected on 3T General Electric 750 scanners at two institutes (University of Wisconsin-Madison and Medical College of Wisconsin). T1-weighted MRI were acquired using MPRAGE (magnetization prepared gradient echo sequence, TR/TE = 604 ms/2.516 ms, TI = 1060.0 ms, flip angle = 8°, FOV = 25.6 cm, 0.8 mm isotropic) \citep{hwang.2020}. Resting-state functional MRI (rs-fMRI) were collected using SMS (simultaneous
multi-slice) imaging \citep{moeller.2010} (8 bands, 72 slices, TR/
TE = 802 ms/33.5 ms, flip angle = 50°, matrix = 104 . 104,
FOV = 20.8 cm, voxel size 2.0 mm isotropic) and a Nova 32-channel
receive head coil. The participants were asked to fixate on a white cross at the center of a black screen during the scans \citep{patriat.2013}. 

MRIs were processed following the Human Connectome Project (HCP) minimal processing pipelines, which are based on FreeSurfer and FSL \citep{glasser.2013}. The T1-weighted images underwent non-linear registration to MNI space, segmentation into structures, reconstruction of cortical surfaces, and surface registration using the "fsaverage" template. Resting-state fMRI (rs-fMRI) images were corrected for spatial distortions using spin echo unwarping maps, realigned to account for subject motion, registered to structural images, bias field corrected, normalized to a global mean, masked, and mapped to native cortical surface space
\citep{garcia.2022}. Additional preprocessing on rs-fMRI images was conducted using AFNI, which included motion regression using 12 parameters, removal of signal changes in white matter and cerebrospinal fluid, global signal regression, and band-pass filtering (0.01–0.1 Hz) \citep{cox.1996, hwang.2020}.

For connectivity analysis, 360 regions from the Glasser parcellation and 19 subcortical regions from FreeSurfer were used \citep{glasser.2016, fischl.2002}. 
In building the Pearson correlation over a sliding window, \citet{shirer.2012} and \citet{leonardi.2015} reported that brain states could be  identified with a window size in the range of 30–60 seconds. \citet{allen.2014} suggested a window size of 44 seconds. \citet{huang.2020.NM} used 43 seconds for Human Brain Connectome Data. \citet{chung.2024.PLOS} used 40 seconds for Wisconsin Twin study. In our study, we used two different window size: 20 TRs = 16.04 seconds and 50 TRs = 40.1 seconds. We used up to 1444 time points that are common across all subjects. There are a total of $T$=1444-20+1 = 1425 sliding windows for 20 TRs and 
$T$=1444-50+1 = 1395 sliding windows for 50 TRs. The 180 cortical regions in Glasser's parcellation are indexed from 1 to 180 for the left hemisphere and 181 to 360 for the right hemisphere, while the 19 subcortical regions are indexed from 361 to 379. Thus we have 1425 time varying correlation matrices of size  379 $\times$ 379 for each subject as the input to the analyses performed in this study.

\subsection{Interpretation of Topological Phase Diagram}
\label{sec:TPD}

Based on the proposed method, we embedded time varying brain network into   Topological Phase Diagrams (TPD) for sliding window of size 20TRs. We have performed the Gaussian kernel smoothing on  TPD  to obtain their empirical probability distributions, which are smooth enough to guarantee the application of the random field theory. Kernel bandwidth $\sigma =0.01$ is used. We then computed the mean and standard deviation of TPD within each group (Figure \ref{fig:TPDstat1}-left).

HC exhibits slightly {\em wider} cumulative birth values compared to individuals with TLE, this suggests that in HC, new components (0D topology) appear across a broader range of filtration values. This indicates a more diverse and possibly more dynamic change of modular structure in HC over time, where various components emerge at different stages of graph filtration. The wider range of birth values in HC could reflect a more heterogeneous or adaptable brain network, where different regions or modules become connected at various thresholds, possibly allowing for more dynamic functional integration and segregation over time.

A death value indicates the scale at which a loop disappear  as the filtration parameter increases. TLE patients show {\em wider} cumulative death values, this implies that the cycles (1D topology) in their brain networks persist across a broader range of filtration values before disappearing. This suggests that the brain networks of TLE under goes more rapid topological changes in 1D topology compared to HC over time. These observations could be related to the underlying pathology of TLE, where altered neural connectivity or brain network disruptions are common. These findings align with  static graphic theory metrics on rs-fMRI that demonstrate global changes in connectivity like a decreased clustering coefficient and increased rich club proportion, both suggestive that networks in TLE are less integrated into the global brain networks with a more rigid structure \citep{struck.2021, mazrooyisebdani.2020, liao.2010,lopes.2017}. In our previous study on static rs-fMRI brain networks \citep{chung.2023.NI}, we found TLE is far more sparse compared to more densely connected HC. This could be due to the brain's adaptation to chronic epilepsy, resulting in a network that tend to fluctuate more in its loops or cycles to maintain connections.

We formally  tested the mean smoothed TPD differences shown in Figure \ref{fig:TPDstat1} using the random field theory in a small square domain of size 0.1 containing the signal regions. The $t$-random field (HL - TLE) with 50+101-2 =  149 degrees of freedom was constructed (Figure \ref{fig:TPDstat1}-right). The maximum $t$-statistic is 4.10 (significant at 0.0057) while the minimum $t$-statistic is -2.66 (not significant). The multiple comparisons correction was done through the proposed random field theory. Only the red regions are statistically significant. In summary, the presence of more dynamically changing features (0D topology) in HC could imply a more dynamically reconfigurable and adaptable network, where subnetworks are made connected and dissolved as necessary. For TLE patients, the persistence of 1D topological features (which connected nodes might indicate a more rigid or less adaptable network structure, potentially reflecting pathological or compensatory mechanisms.

We repeated the analysis with larger sliding window size of 50TRs with smaller kernel bandwidth $\sigma =0.005$ in a small square domain of size 0.1 containing the signal regions (Figure \ref{fig:TPDstat2}). The use of a larger sliding window provides a broader temporal context, potentially capturing more stable and persistent patterns in the brain's connectivity over time. The subtle distributional differences observed in smaller temporal resolution of 20TRs disappeared. Still we detected statistically significant TPD differences. The maximum $t$-statistic is 4.59 (significant at 0.0036) while the minimum $t$-statistic is -4.57 (significance at 0.0039). The multiple comparisons correction was done through the proposed random field theory. The maximum cumulative birth and death values occur at (0.8778, 0.0998) for HC and at (0.8725, 0.0998) for TLE, suggesting a subtle difference in the persistence of connected components between the groups. The lower maximum birth for TLE indicate a tendency for their brain networks to maintain connected components (0D topology) for a longer duration when observed at the larger temporal resolution. Conversely, the close alignment in the maximum accumulated death values for both groups suggests that the 1D topological features, such as cycles and loops, exhibit a similar degree of persistence between HC and TLE over larger temporal resolution (50TRs) but more rapid changes for TLE over shorter temporal resolution (20TRs).

The mean maps presented in Figures \ref{fig:TPDstat1} and  \ref{fig:TPDstat2} demonstrate the robust nature of TPD as a feature, with both healthy controls (HC) and temporal lobe epilepsy (TLE) patients showing similar distribution of average cumulative birth and death values. This similarity underscores the stability of the topological features captured by TPD, regardless of the underlying condition.

\begin{figure}[t]
\begin{center}
\includegraphics[width=1\linewidth]{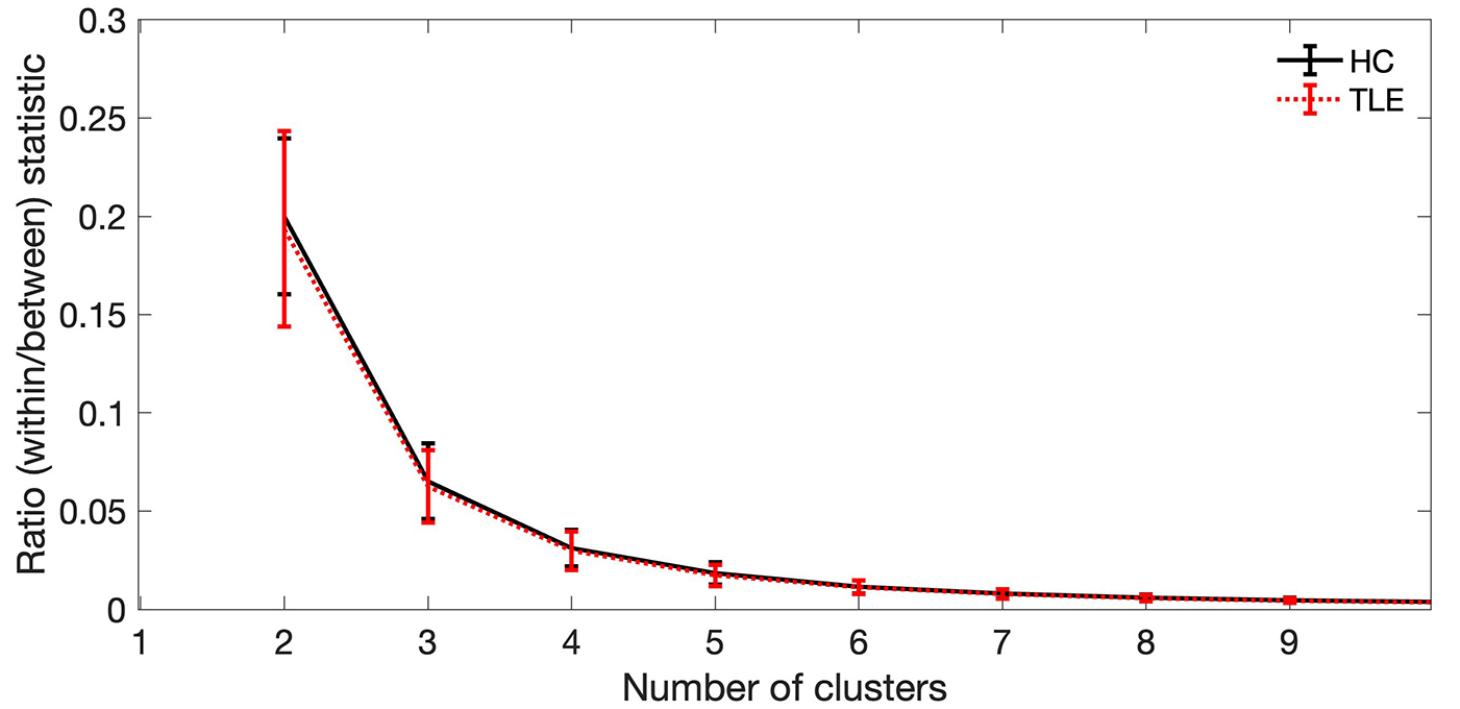}
\caption{Ratio of within over between cluster distance for window size 20 TRs. The point of most rapid change ($k$=3) of ratio determines the optimal number of cluster in the elbow method. The ratio is stable across group labels.}
\label{fig:ratio}
\end{center}
\end{figure}

\begin{figure}[t]
\begin{center}
\includegraphics[width=1\linewidth]{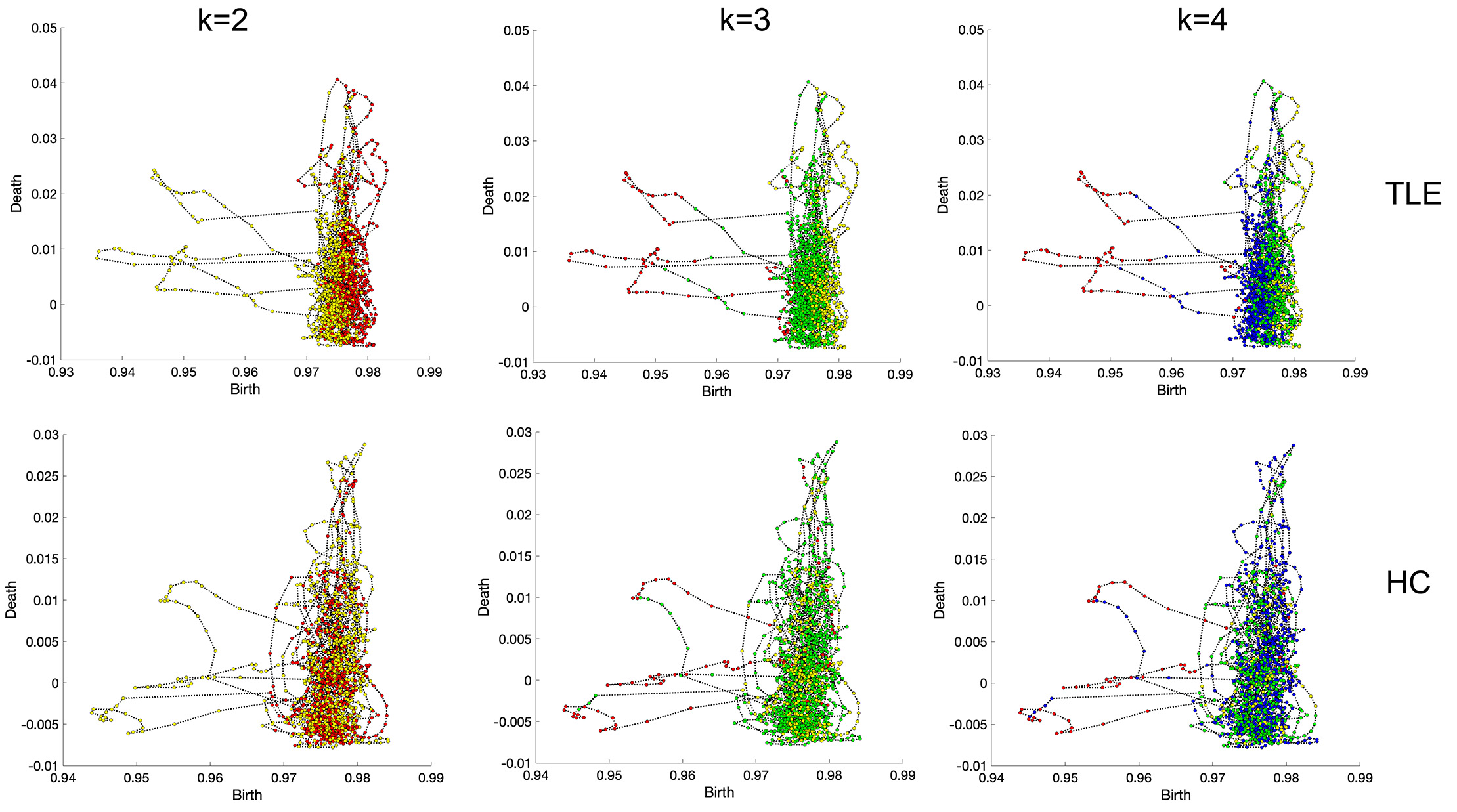}
\caption{Topological phase diagram (TPD) illustrating the temporal evolution of functional brain network topology for two representative subjects for the window size of 20 TRs (16 seconds). Despite the lack of synchronization in rs-fMRI connectivity across individuals, the phase diagram reveals a consistent pattern. Topological clustering identifies transient states, each represented by a distinct color. We determined that $k=3$ is the optimal number of clusters to represent these states.
The threds-like patterns observed to the left of the clusters are likely to be topological noise, which is not observed with longer window sizes.}
\label{fig:TPD-coord1}
\end{center}
\end{figure}

\subsection{Topological transient state space}
\label{subsec:states}

The application of random field theory to TPD revealed two distinct signal clusters in the $t$-field for window size 20 TRs. By performing topological clustering, we will confirm that these signal regions correspond to two unique transient states within the dynamics of resting-state fMRI connectivity (top middle in \label{fig:TPD}). This relation between the statistical findings and the topological clustering underscores the effectiveness of our approach in capturing meaningful patterns in the temporal evolution of brain networks.

For \(p\)=379 brain regions, we estimated \(p \times p\) dynamically changing correlation matrices \(C(t)\) for each subject over sliding windows of size 20. Let \(\mathbf{C}(t)\) represent the vectorization of the upper triangle of  \(C(t)\) at time  \(t\) into a  vector of size \(p(p+1)/2\). The collection of \(\mathbf{C}(t)\) over all timepoints is then fed into topological clustering to identify the recurring brain connectivity states in each subject.  This results in the time series of sorted birth and death sets $B(t)$ and $D(t)$. Every birth and death sets over all time points are then feed into topological clustering. We cluster individual brain networks without imposing any group-level constraints \citep{chung.2024.PLOS}.  After clustering, each correlation network \(C(t)\) is assigned a discrete state indexed with  integer between 1 and \(k\). These discrete states serve as the foundation for investigating the dynamic patterns of brain connectivity \citep{ting.2018}. To ensure convergence, both topological clustering and \(k\)-means clustering were repeated 10 times with different initial centroids, and the best result (the one with the smallest within-cluster distance) is reported.

For each value of $k$, we computed the ratio of the within-cluster  to between-cluster distances. The ratio assesses the goodness-of-fit of the cluster model.  The optimal number of clusters \(k\) was determined using the elbow method \citep{allen.2014,rashid.2014,ting.2018,huang.2020.NM}. The elbow method identifies the point where there is the most significant change in the ratio's slope, which corresponds to \(k=3\) in our study  (Figure \ref{fig:ratio}). Compared to standard \(k\)-means clustering, topological clustering achieves up to a six-fold reduction in the ratio, indicating a superior model fit over \(k\)-means \citep{chung.2024.PLOS}. Figure \ref{fig:ratio} illustrates the results of this ratio comparison over different $k$. The TPD is color-coded according to cluster labels, revealing two dominating states, while the third state is likely a noisy intermediate state. Figures \ref{fig:TPD-coord1} and Figure \ref{fig:TPD-coord2} displays the result for 20 and 50 TRs respectively.  Once the network reaches a state, it tends to remain in the same state for an extended period resulting in the observed clustering pattern.

\begin{figure}[t]
\begin{center}
\includegraphics[width=1\linewidth]{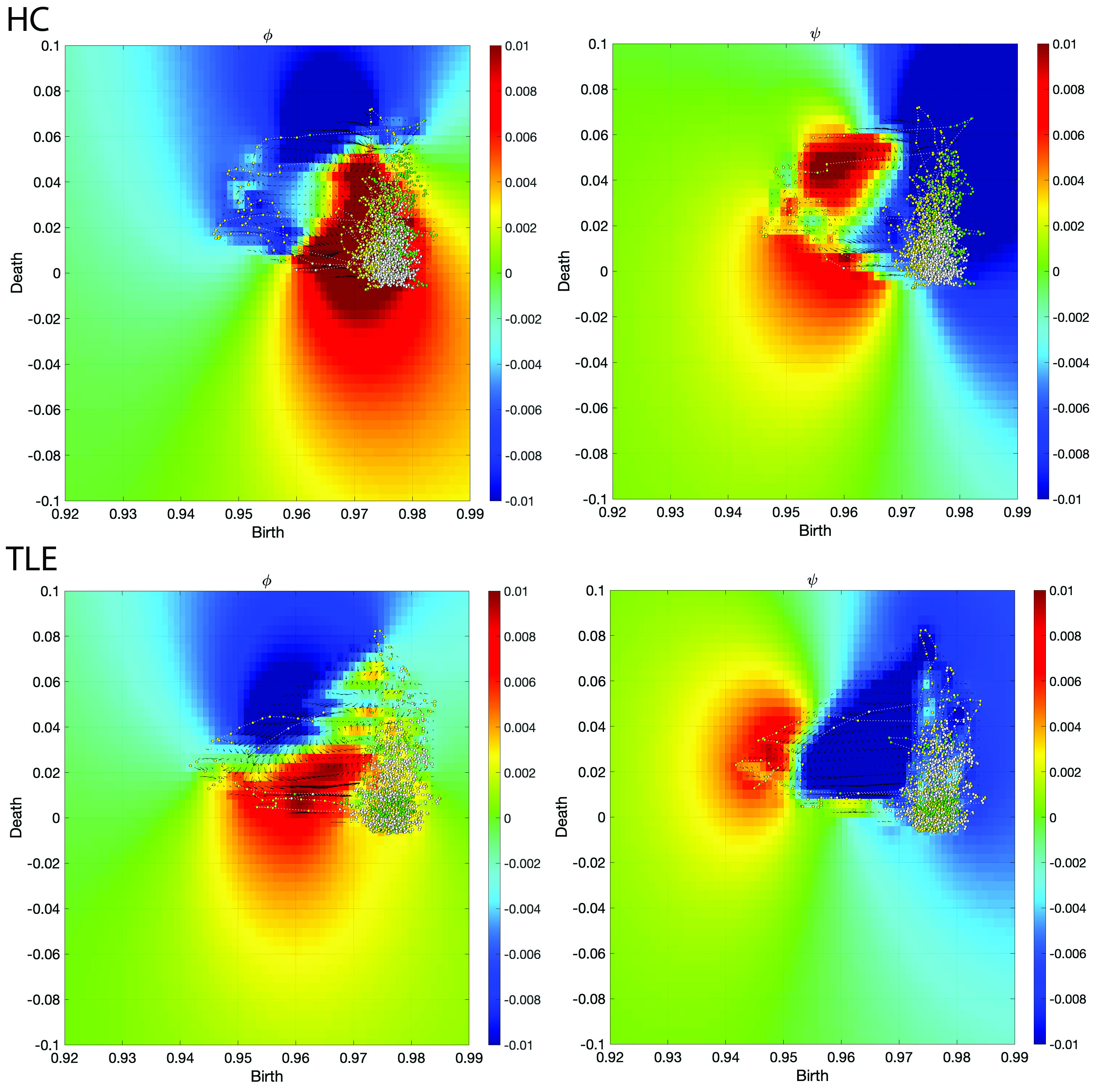}
\caption{The estimated scalar functions \(\phi\) and \(\psi\) for two representative subjects, for window size 20 TRs, are shown. Colored dots represent rs-fMRI brain networks at each sliding window, totaling 1425 networks per subject. Arrows denote the topological gradient \(w_k\) derived from TPD. }
\label{fig:hodge-6subjects}
\end{center}
\end{figure}

\subsection{Hodge decomposition}

\begin{figure}[t]
\begin{center}
\includegraphics[width=1\linewidth]{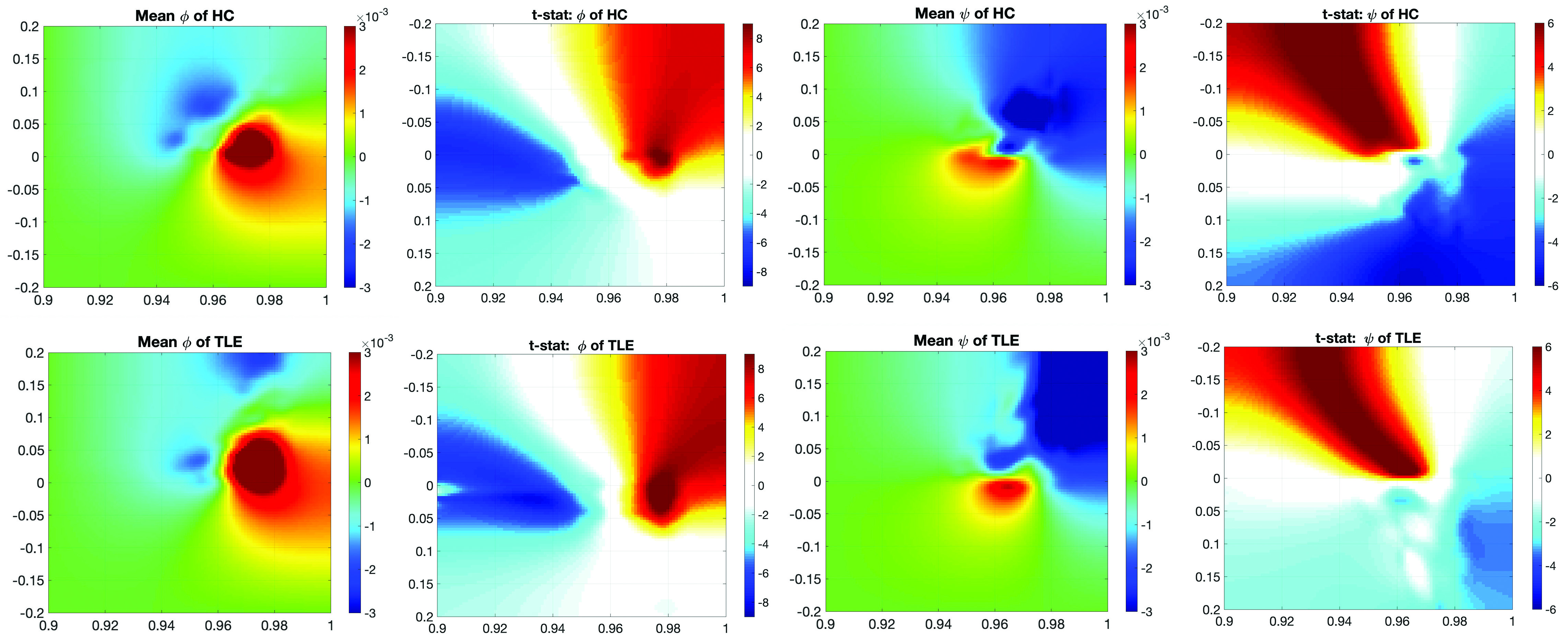}
\caption{The mean estimated scalar functions \(\phi\) and \(\psi\) for both the HC and TLE groups, computed over sliding windows of size 20 TRs. A consistent within-group pattern is observed, characterized by two distinct modes of positive and negative domains in the scalar fields \(\phi\) and \(\psi\). The one-sample $t$-statistic in each group (2nd and 4th columns) shows significant, consistent population-level signals detected mainly in two clearly separable regions, indicating the existence of two major topological states. However, no statistically significant regions are detected through the two-sample $t$-statistic between the mean scalar fields of the groups.}
\label{fig:hodge-means}
\end{center}
\end{figure}

We further explored if the Hodge decomposition can be used to quantify the direction or flow information in TPD. Figure \ref{fig:hodge-6subjects} displays  the estimated scalar fields \(\phi\) and \(\psi\) for two regenerative subjects for smaller window of size 20 TRs. Even though the patterns of scalar fields \(\phi\) and \(\psi\) are varied, they show consist pattern of one dominant positive and one dominant negative domains that split TPD. All three states (colored differently) are mostly observed in regions of positive \(\phi\), while regions of negative \(\psi\) might correspond to transitional or less active brain states. Thus, we computed the mean scalar fields within each group (Figure \ref{fig:hodge-means}) and tested for statistical significance. We observed consistent population specific patterns having two distinct modes of positive and negative domains in the scalar fields in each group. The one-sample $t$-statistic in each group (2nd and 4th columns) shows significant, consistent population-level signals detected mainly in two clearly separable regions, indicating the existence of two major topological states. However, in the two sample $t$-test comparing the groups, we could not find any region in TPD that is statistically significant at 0.01 level after the multiple comparison corrections using the random field theory.

\subsection{Time-frequency analysis of topological phase diagram}

To examine the dynamical behavior of topological features over time, we applied a time-frequency analysis to the evolution of cumulative birth and death values using the Short-Time Fourier Transform (STFT) \cite{durak.2003}. This method partitions the time series data into shorter segments of equal length and computes the Fourier Transform separately on each segment. This approach captures both temporal and frequency information, allowing us to analyze the changes in network topology across different scales and moments using the TPD trajectories. The STFT was computed from the time series of  both the birth and death sets. The STFT is mathematically represented as:
\[
X(\tau, \omega) = \int x_t w(t-\tau) e^{-j \omega t} dt
\]
where $x_t$ is the time series of the normalized cumulative birth values used in plotting the $x$-coordinates of TPD in (\ref{eq:x_t}). \(w(t)\) is the window function, \(\tau\) represents the time shift, and \(\omega\) is the angular frequency.
We can compute STFT similarly for death values using $y_t$ in (\ref{eq:y_t}). The power spectral density (PSD) is then calculated from the STFT to quantify the power of frequencies present in the birth and death sequences over time:
\[
P(\tau, \omega) = |X(\tau, \omega)|^2
\]
The PSD is subsequently converted to a logarithmic scale (decibels, dB) to enhance the dynamic range and visual interpretation:
\[
P_{dB}(\tau, \omega) = 10 \log_{10} P(\tau, \omega).
\]

\begin{figure}[t]
\begin{center}
\includegraphics[width=1\linewidth]{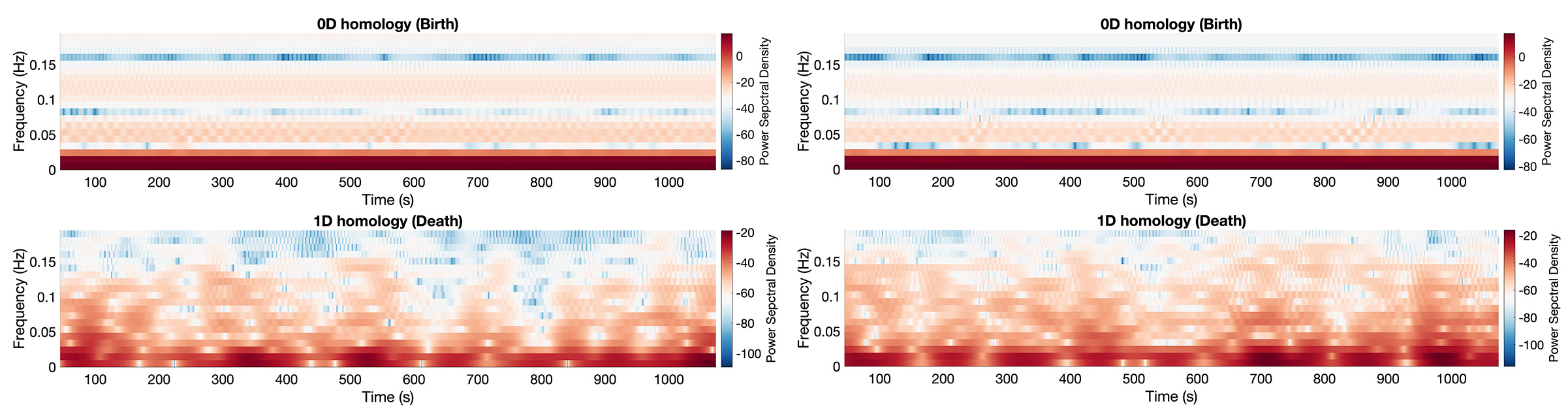}
\caption{Spectrogram for a representative HC (left) and TLE (right) subject. The power spectral density (PSD) levels indicate the intensity of topological signals across the frequency spectrum. The consistently high PSD in the low-frequency range (0-0.03 Hz) for 0D homology suggests a persistent presence of a tree-like backbone (maximum spanning tree), which remains stable throughout the observation period. In contrast, the 1D homology exhibits more variable signals, corresponding to the dynamic formation and dissolution of cycles as nodes within the tree connect or disconnect. This variability underscores the transient nature of higher-dimensional topological structures in the brain networks.}
\label{fig:power-subjects}
\end{center}
\end{figure}

\begin{figure}[t!]
\begin{center}
\includegraphics[width=1\linewidth]{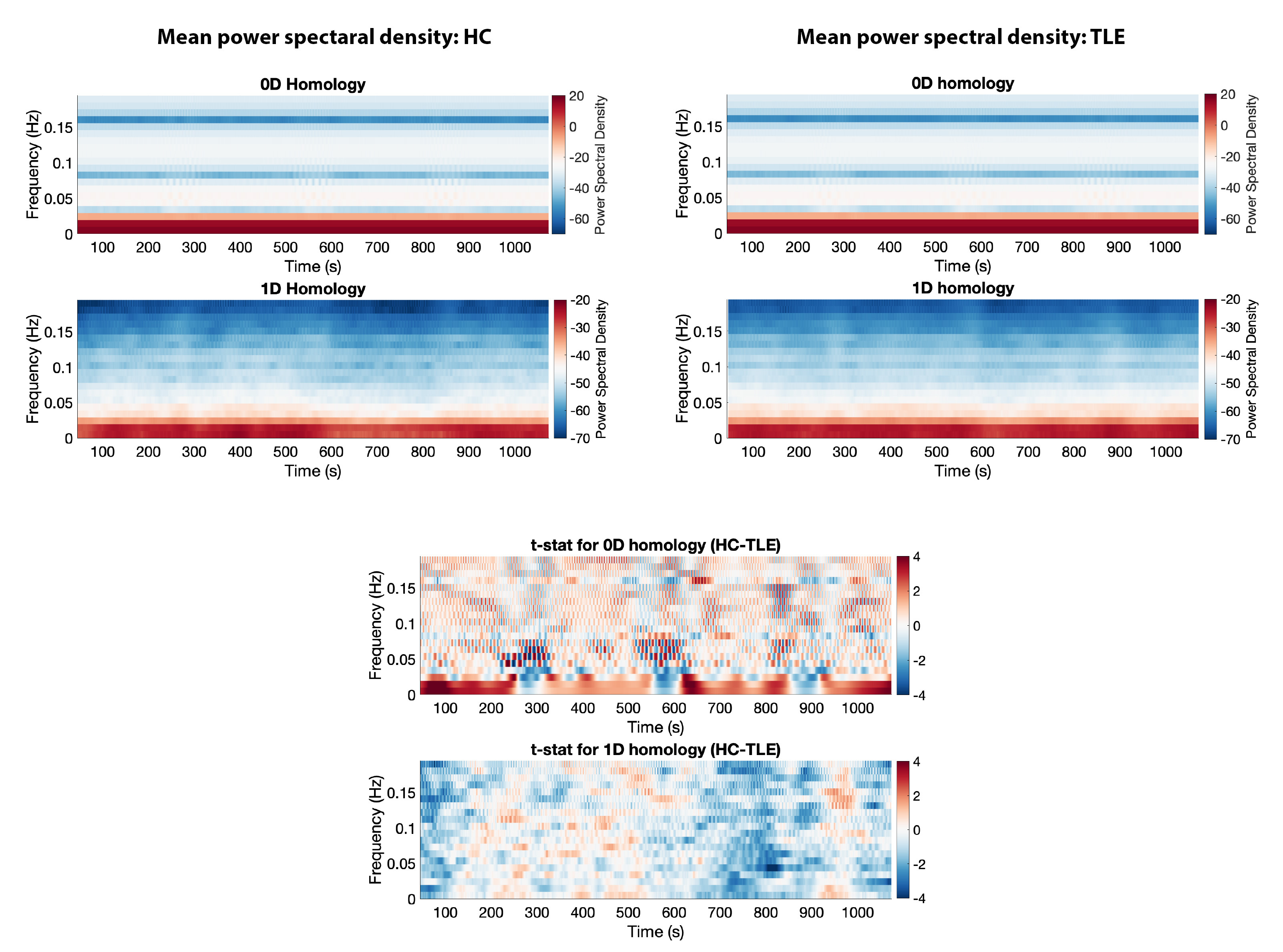}
\caption{Mean spectrogram for HC (top left) and TLE (top right) groups. The consistently high PSD in the low-frequency range (0-0.02 Hz) for 0D homology suggests a persistent presence of a tree-like backbone (maximum spanning tree), which remains stable throughout time in both groups. In contrast, the 1D homology exhibits more variable signals, corresponding to the dynamic formation and dissolution of cycles as nodes within the tree connect or disconnect. Thus the mean PSD of 1D homology is weaker in the low-frequency range compared to 0D homology. 
Bottom: Two sample $t$-statistic at each time and frequency. HC shows stronger PDS in 0D homology indicating the backbone of normal control is more stable and consistent over time. TLE shows stronger PDS in 1D homology indicating more dynamic changes of connections that form or destroys cycles.}
\label{fig:power-means}
\end{center}
\end{figure}

For the time-frequency analysis, we used time varying correlation matrix obtained from the sliding window of size 50TRs.
We chose parameters for the STFT that balance time and frequency resolution, which are critical for capturing the relevant dynamics in the data. The sampling rate given as \(1 / 0.802 = 1.247 \) Hz, corresponding to the inverse of the sampling interval of the rs-fMRI data. This rate defines the temporal granularity at which the Fourier Transform is applied. We used the window size set to 10\% of the total duration of the time series of birth and death values. This window size is selected to provide a reasonable trade-off between time and frequency resolution, allowing us to detect changes over a meaningful time scale without blurring important features. Then we sliced the window one time step at a time 
 ensuring high resolution in the time-frequency representation by providing a smooth transition between consecutive windows.
 The results are visualized using spectrograms that plot the time-varying frequency content of  TPD for two representative subjects (Figure \ref{fig:power-subjects}).

We conducted a group-level statistical analysis on the spectrograms. Traditionally, it has been challenging to directly compare time-varying rs-fMRI connectivity matrices at identical time points across subjects due to lack of synchronization in neural activity patterns. Our approach addresses this challenge by focusing on the underlying topological structures rather than precise temporal alignment, allowing for a comparison that bypasses synchronization issues. We computed group mean spectrograms (Figure \ref{fig:power-means}-top). For 0D topology, the frequency bands between 0-0.02Hz exhibited highly consistent activity across different subjects, suggesting common stable patterns of maximum spanning tree over time. These stable frequencies likely correspond to fundamental underlying backbone of brain network that are preserved across individuals and over time. For 1D topology, the consistently observed frequency bands between 0-0.02Hz correspond to the birth and death of cycles, which are formed when nodes in the underlying 0D structure (tree) are connected or disconnected. These transitions manifest as the transient topological changes observed in our previous state space modeling.

The statistical significance of differences in group means was determined using a two-sample $t$-test under the assumption of equal variances. Given the discrete sampling over the frequency domain, we could not apply the random field theory. Instead, we utilized the online permutation testing to assess significance \citep{chung.2019.CNI}. Any regions showing dark blue or dark red are statistically significant below the 0.01 level.  There were numerous statistically significant time points in the frequency range 0-0.02Hz for 0D homology (underlying MST), indicating that the dynamics of 0D homology is significant  discriminating factor between groups. Still the underlying backbone of brain network is fairly stable across subjects and groups in the temporal resolution of 50 seconds (corresponding to 0.02Hz). Much higher PSD for HC implies that brain networks of HC is more dynamically changing in the underlying MST compared to more rigid backbone of TLE which is less changing over time. 

In contrast, the group difference in 1D homology is not as significant as 0D homology. 
However, we still observed that PSD is stronger for TLE in the 0-0.02Hz range for 1D homology, indicating more frequent and pronounced topological transformations in TLE patients compared to HC. The brain network of TLE is more dynamically connecting and disconnecting edges between the nodes of the MST that create or destroy cycles. This suggests that TLE is associated with more dynamic alterations in network cycles, potentially reflecting compensatory mechanisms or heightened neural plasticity in response to rigidity in the underlying backbone of brain network.

\subsubsection{Relating general intelligence to topology}

\begin{figure}[t]
\begin{center}
\includegraphics[width=1\linewidth]{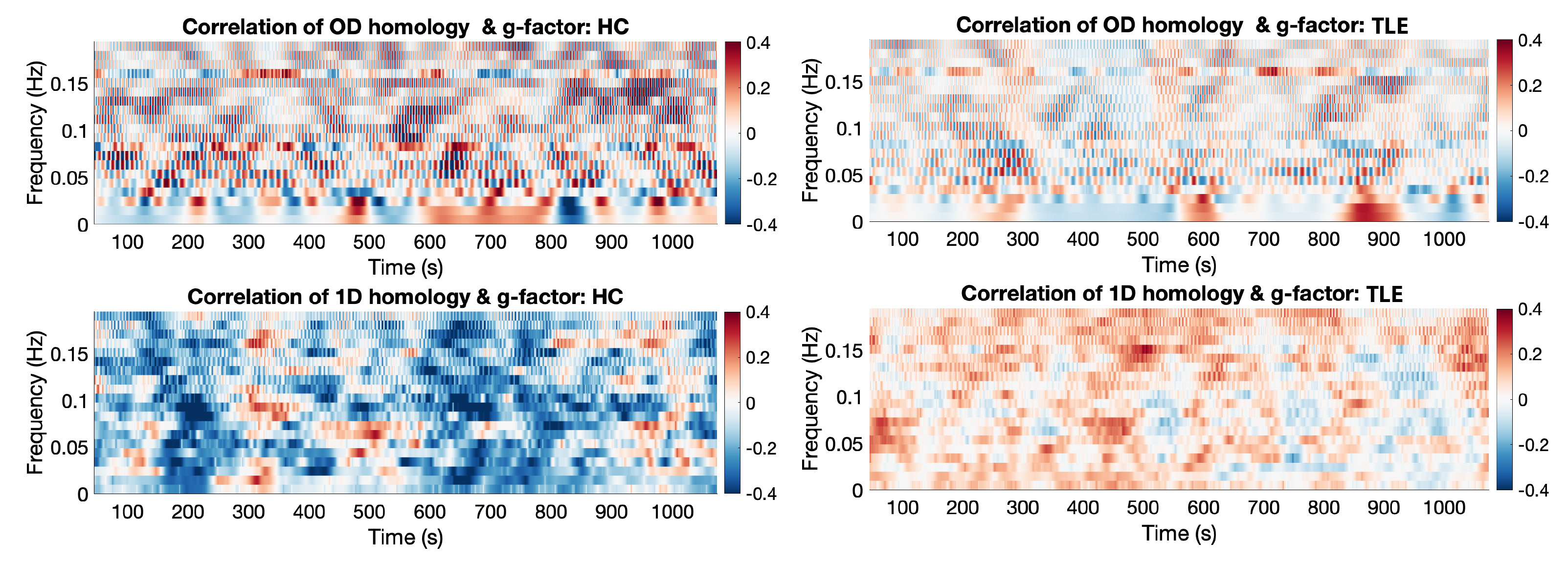}
\caption{Correlation of power spectral density (PSD) with g-factor across 0D and 1D homological features at each time and frequency. The 0D homology correlations with the g-factor do not significantly differentiate between the groups. In contrast, the 1D homology correlations with the g-factor show significant differences between the groups. For healthy controls (HC), the 1D topology is predominantly negatively correlated with the g-factor, suggesting that fewer dynamic changes in connectivity are associated with higher cognitive function. This implies that a more stable maximum spanning tree (MST) suffices for efficient information transfer in HC brains. On the other hand, in temporal lobe epilepsy (TLE) subjects, the 1D topology is mostly positively correlated with the g-factor, indicating that higher cognitive function is associated with increased dynamic connectivity, possibly as a compensatory mechanism for a less flexible backbone network.}
\label{fig:power-corr}
\end{center}
\end{figure}

Temporal lobe epilepsy (TLE) is a neurological condition where seizures primarily originate in the temporal lobes of the brain, which are crucial for processing sensory input and are important in memory, speech, and visual recognition processing \citep{khalife.2022,rzezak.2017}. Thus, TLE can affect various cognitive functions. Since the temporal lobes play a key role in memory formation and retrieval, individuals with TLE often experience memory problems. The left temporal lobe is particularly important for language. Damage or disruption in this area due to epilepsy can affect language comprehension and production, skills that are often measured in IQ tests. While the frontal lobes are more commonly associated with executive functions, seizures in the temporal lobe can also impact these higher-order cognitive processes due to the interconnectedness of neural pathways. TLE can influence mood and behavior, which indirectly might affect cognitive performance. Depression and anxiety are common in individuals with epilepsy and can impair cognitive function and overall performance on tasks that measure aspects of the g-factor. 

Thus, we investigated if {\em g-factor} is related to the over all dynamics of the topological changes in brain networks \citep{johnson.2008}. The general intelligence or g-factor represents  the common skills and cognitive abilities that underpin a wide range of intellectual tasks. General intelligence is thought to reflect a common cognitive basis for all mental activities. For instance, individuals who perform well on a verbal test are also likely to perform well on other cognitive tasks, suggesting the presence of a general intelligence that affects all cognitive abilities. The g factor is a strong predictor of academic and occupational success, more so than any specific ability. It's used in educational and psychological settings to help predict performance across a range of activities and environments. Factor analysis was used to examine the data from multiple cognitive tests like the WASI Vocabulary and Block Design tests among others \citep{irby.2013}, to determine if they can be explained largely or entirely by a single common factor, which would be evidence supporting the existence of {\em g}. By standardizing these scores (mean of 0, standard deviation 1), we ensured that the results are normalized, making it easier to compare individuals regardless of the original scale of the measures used.

We analyzed the correlation between the power spectral density (PSD) at each time and frequency with the g-factor in two groups: 34 HC and 94 subjects with TLE. The results, depicted in Figure \ref{fig:power-corr}, show correlations thresholded at $\pm 0.4$ for clearer visualization. We used the correlation differences between the groups in determining statistical significance between the groups using the online permutation test \citep{chung.2019.CNI}. It is unclear if the correlation of 0D topology with the g-factor is discriminative enough between the groups. However, the correlation of 1D topology with the g-factor shows significant group differences and trends. For the HC group, the correlation between the 1D topological features and the g-factor is predominantly negative. This suggests that higher cognitive ability is associated with fewer dynamic changes in 1D topology. Specifically, a more intelligent brain may not necessarily exhibit increased connectivity between the nodes in the backbone network. This implies that the maximal spanning tree (MST) of the brain network is efficient in information transmission, and additional connections beyond this backbone may be redundant and unnecessary. Conversely, in the TLE group, the 1D topology shows a predominantly positive correlation with the g-factor. This indicates that for individuals with TLE, higher cognitive function is associated with more frequent changes in connectivity. The backbone of the TLE brain network is less dynamic and more rigid, which may necessitate increased connectivity and disconnection  to compensate for the insufficiencies in the backbone structure.

In summary, the correlation of 1D topology with the g-factor differentiates between HC and TLE groups. For HC, cognitive efficiency seems to rely on a stable backbone network with minimal additional connections, whereas in TLE, cognitive function appears to be supported by a more dynamic and compensatory connectivity pattern. This highlights the contrasting network dynamics in HC and TLE, with potential implications for understanding the neural mechanisms underlying cognitive function in these populations.

\begin{figure}[t]
\centering
\includegraphics[width=1\linewidth]{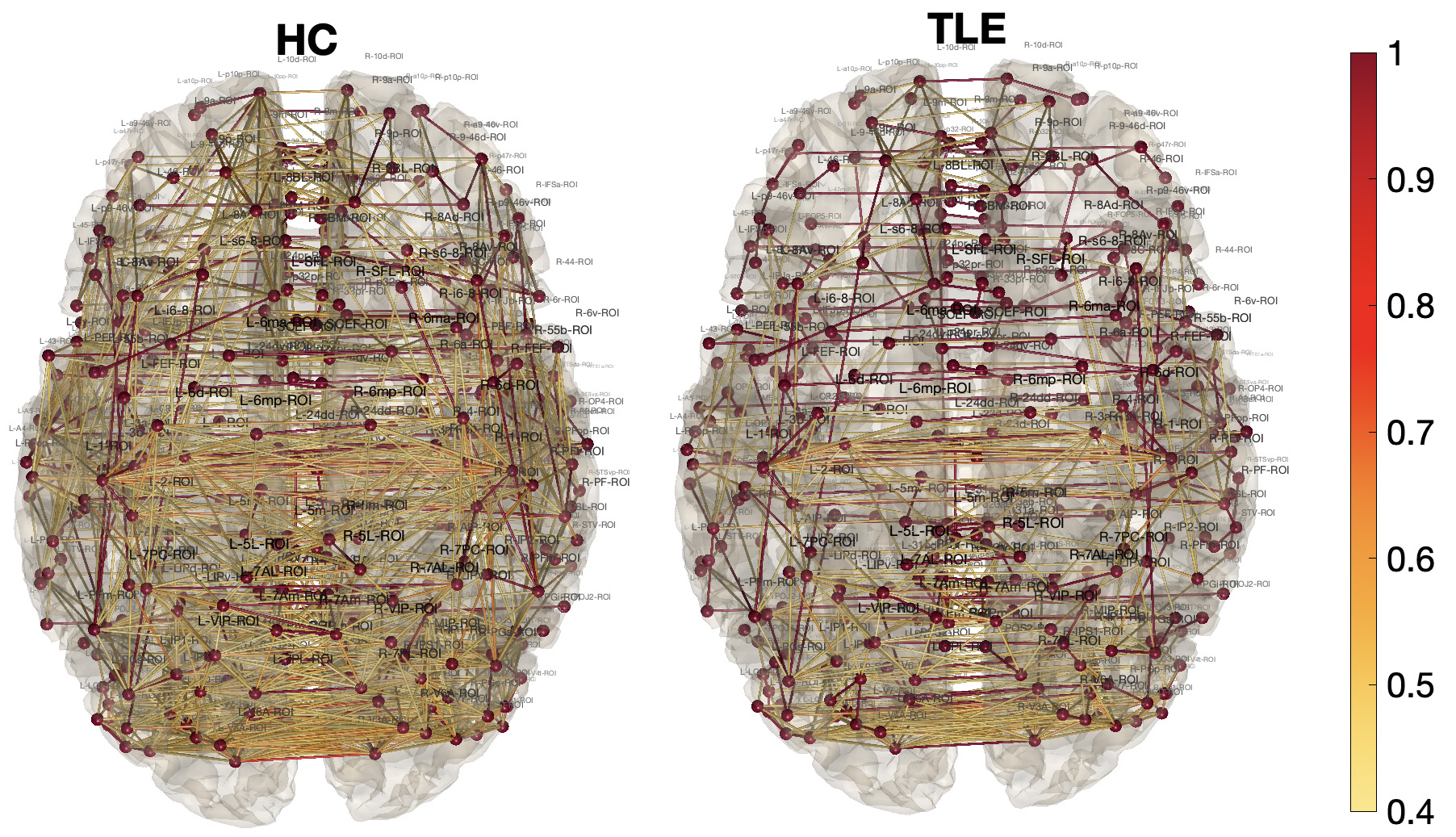}
\caption{Average correlation brain networks for 50 healthy controls (HC) and 101 temporal lobe epilepsy (TLE) patients, overlaid on the gray matter boundary of the MNI template. The network of TLE patients appears notably sparser than that of HCs. This static summary significantly simplifies the data, potentially overlooking more complex differential signals that manifest over time. Therefore, we have partitioned the network into 0D and 1D homologies. Dark red edges represent the maximum spanning tree (MST), which characterizes the 0D homology. The relative sparsity of the TLE network arises not from the 0D homology but rather from the 1D homology, which involves non-MST edges forming cycles. The dynamic nature of the TLE brain network is thus characterized by fluctuations in non-MST edges that significantly differ from those observed in HC.}
\label{fig:correlation}
\end{figure}

\section{Conclusion \& Discussion}

\begin{figure}[t]
\includegraphics[width=1\linewidth]{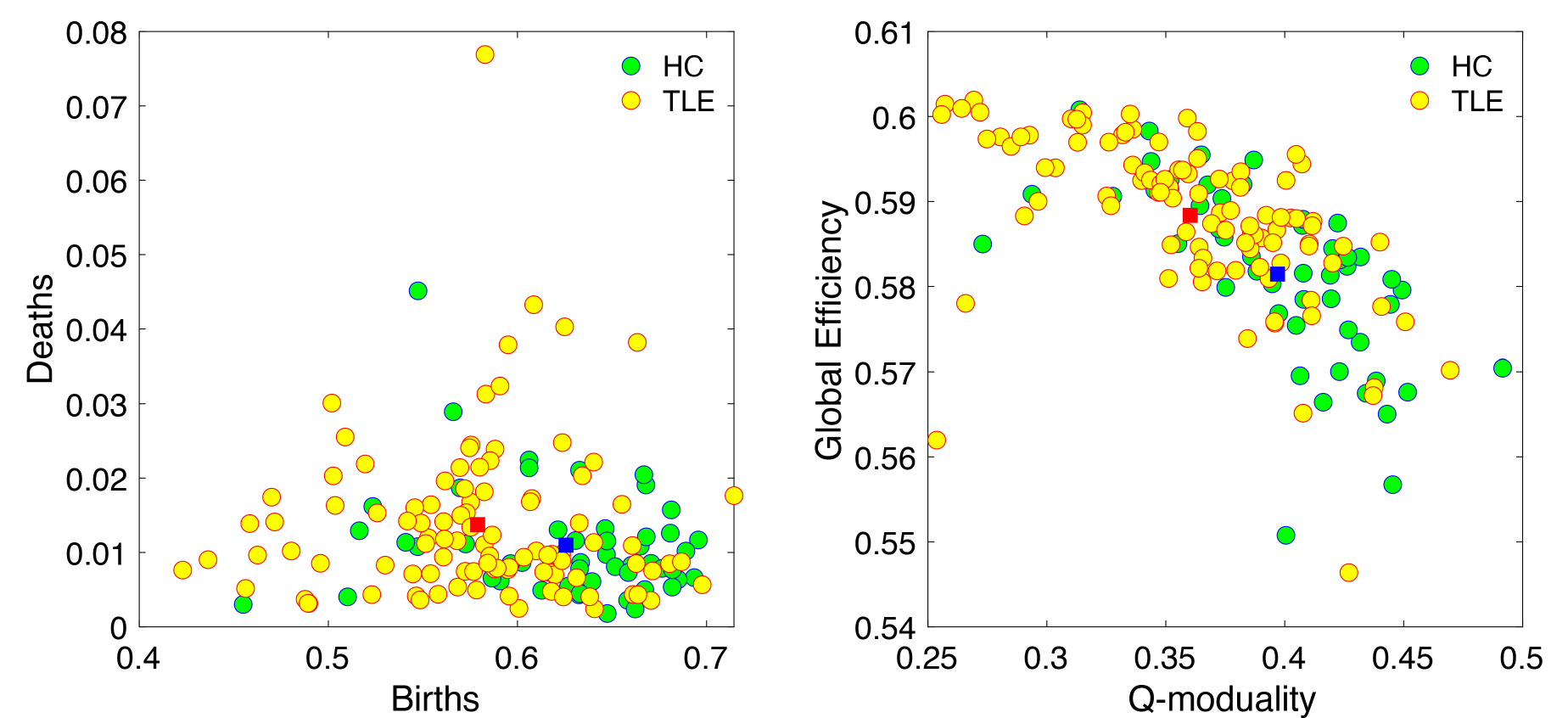}
\caption{Left: Topological phase diagram of 151 subjects using the Pearson correlation matrix per subject as input. Green circles are HC and yellow circles are TLE. The blue square is the topological mean of HC while the red square is the topological mean of TLE. The horizontal axis represents 0D topology (connected components) through birth values while the vertical axis represents 1D topology (circles) through death values.
The embedding shows that the topological separation in static networks is mainly through 0D topology. Right: Embedding through graph theory features global efficiency at vertical and Q-modularity at horizontal axes.}
\label{fig:embedding}
\end{figure}

In our previous study focusing on the static summary of brain networks at rest \citep{chung.2023.NI}, we observed that networks associated with temporal lobe epilepsy (TLE) were sparser compared to those of healthy controls (HC) (Fig \ref{fig:correlation}). In that study, we calculated pairwise Pearson correlations between brain regions over all time points for each subject, using static summary connectivity to quantify group differences between TLE and HC. However, this approach, which collapses the entire time series into a single connectivity matrix, may oversimplify the data, failing to capture potential dynamic signals present throughout the time points. If we apply the proposed \textit{Topological Phase Diagram} (TPD) to static Pearson correlation matrices, we obtain Figure \ref{fig:embedding}. Here, the topological group means (represented by a blue square for HC and a red square for TLE) exhibit significant separation in 0D topology. This simplification correlates well with existing graph theory features such as global efficiency and Q-modularity, yet it does not provide much insight into the dynamics of network changes \citep{garcia.2022,newman.2004}.

In our current method, we decompose the entire time series of networks into disjoint 0D and 1D topological components. This decomposition facilitates the embedding of the entire evolution of brain networks into a visually interpretable 2D plane via the \textit{Topological Phase Diagram} (TPD). Through various analyses conducted on the TPD, we have characterized the dynamics of the TLE brain network in contrast to that of HC.

The maximum spanning tree (MST), which characterizes 0D homology, demonstrates remarkable stability in both healthy controls (HC) and temporal lobe epilepsy (TLE) patients over extended periods. This stability suggests that the primary connectivity pathways, as represented by the MST, are only subtly affected by epilepsy. This could indicate that the fundamental organizational principles of brain networks remain resilient despite the alterations associated with TLE. Further investigation is needed to determine if the observed stability of the MST, serving as a backbone for dynamic changes, is related to structural connectivity, which remains consistent over time. However, subtle fluctuations in the MST can still differentiate between the groups within shorter periods. HC exhibits more dynamic changes in the underlying backbone network compared to TLE. TLE seems more rigid and less flexible in the backbone network.

The group differences are primarily characterized by dynamic changes in edges that are not part of the MST (0D topology). The connection or disconnection of these edges leads to the creation or destruction of cycles, thereby altering the 1D topology we observe. The cyclic changes that differentiate the groups are more pronounced in longer periods of time, but are less distinct at shorter time scales. These complex dynamic changes are driving the topological state changes we observe. Brain networks of TLE seems fluctuate more in 1D topology as possible compensation mechanism for less fluctuations in 0D topology.

In healthy individuals with higher cognitive ability, the brain exhibits less fluctuation in 1D topology. The existing backbone network, represented by the MST, appears sufficient to convey all necessary information, reducing the need for additional redundant connections through 1D topology. This contrasts with findings in individuals with temporal lobe epilepsy (TLE), where the brain shows greater fluctuation in 1D topology. TLE seems to affect the underlying backbone network (0D topology), making it less dynamic and more rigid. To compensate for the limited information flow in such a constrained backbone network, TLE appears to introduce more dynamic fluctuations in 1D topology, creating redundant connections. Further investigation is needed to understand these divergent dynamics, which seem to work in opposite directions in TLE.

In our exploration of the Hodge decomposition, we noticed that the two prominent states predominantly reside within the positive \(\phi\) and negative \(\psi\) domains. This observation offers an intriguing avenue for refining state space models by incorporating constraints based on these domains. The transition probability $P$ from brain network $\mathcal{X}_t$ having state $i$ to brain network $\mathcal{X}_{t+1}$ having state $j$  can be influenced by the scalar fields \(\phi\) and \(\psi\) as follows:
\[
P(X_{t+1} = j | X_t = i, \phi, \psi) = 
\begin{cases} 
p_{ij}, & \text{if } \phi > 0 \text{ and } \psi < 0, \\
0, & \text{otherwise},
\end{cases}
\]
with $\sum_{j=1}^3 p_{ij}=1$. The conditions \(\phi > 0\) and \(\psi  < 0\) act as constraints that can modulate the transition probabilities based on the observed Hodge decompostion. While we did not delve into this aspect in the current study, it presents a compelling direction for future research.

%

\subsection*{Ethics statement}
The study was approved and follows the University of Wisconsin-Madison and Medical College of Wisconsin IRB protocols. The study was conducted ethically following the IRB protocol. 

\section*{Data availability statement}
We made the core MATLAB computer code available at \url{https:// github.com/laplcebeltrami/PH- STAT}. The epilepsy data is not available due to the IRB protocol.

\section*{Acknowledgements}
This study was supported by NIH U01NS093650, NS117568, EB028753,  MH133614 and  NSF MDS-2010778.


\bibliographystyle{plainnat}
\bibliography{reference.2024.05.03}

\end{document}